\documentclass[conference]{IEEEtran}
\usepackage[dvips]{graphicx}
\usepackage[cmex10]{amsmath}
\usepackage{multirow}
\usepackage{epsfig}
\usepackage{mathrsfs}
\usepackage{amssymb}
\usepackage{comment}
\usepackage{bm}
\usepackage{url}
\usepackage{array}
\usepackage{amsthm}
\usepackage{blkarray}
\usepackage{fancyhdr}
\usepackage{enumerate}
\usepackage[lined,boxed,commentsnumbered,ruled,linesnumbered]{algorithm2e}

\renewcommand{\thispagestyle}[1]{} 

\theoremstyle{definition}
\newtheorem{theorem}{Theorem}[section]
\newtheorem{lemma}[theorem]{Lemma}

\newtheorem{definition}{Definition}

\ifCLASSINFOpdf

\else

\fi

\usepackage[tight,footnotesize]{subfigure}

\begin{document}
\pagestyle{fancy}
\IEEEoverridecommandlockouts

\lhead{\textit{Technical Report, Dept. of EEE, Imperial College, London, UK, Jul., 2013.}}
\rhead{} 
%
\title{Partial Network Identifiability: Theorem Proof and Evaluation}
\author{\IEEEauthorblockN{Liang Ma\IEEEauthorrefmark{2}, Ting He\IEEEauthorrefmark{3}, Kin K. Leung\IEEEauthorrefmark{2}, Ananthram Swami\IEEEauthorrefmark{4}, and Don Towsley\IEEEauthorrefmark{1}}
\IEEEauthorblockA{\IEEEauthorrefmark{2}Imperial College, London, UK. Email: \{l.ma10, kin.leung\}@imperial.ac.uk\\
\IEEEauthorrefmark{3}IBM T. J. Watson Research Center, Hawthorne, NY, USA. Email: the@us.ibm.com\\
\IEEEauthorrefmark{4}Army Research Laboratory, Adelphi, MD, USA. Email: ananthram.swami.civ@mail.mil\\
\IEEEauthorrefmark{1}University of Massachusetts, Amherst, MA, USA. Email: towsley@cs.umass.edu
}
}

\maketitle

\IEEEpeerreviewmaketitle

\section{Introduction}
Auxiliary algorithms and selected theorem proofs in \cite{MaInfocom14sub} are presented in detail in this report. We first introduce the auxiliary algorithms in Algorithm DAIL, Determination of All Identifiable Links \cite{MaInfocom14sub}, for identifying special links in triconnected components of Category 2 and 3 in Section II. We then list the theorems in Section III and give the corresponding proofs in Section IV. In the end,
we show some additional simulation results to verify the superior efficiency of GMMP compared with the benchmark solutions. See the original paper \cite{MaInfocom14sub} for terms and definitions. Table~\ref{t notion} summarizes all graph-theoretical notions used in this report (following the convention in \cite{GraphTheory2005}).

\begin{table}[b]
\vspace{.5em}
\renewcommand{\arraystretch}{1.3}
\caption{Notations in Graph Theory} \label{t notion}
\vspace{-.5em}
\centering
\begin{tabular}{r|m{6.26cm}}
  \hline
  \textbf{Symbol} & \textbf{Meaning} \\
  \hline
  $V(\mathcal{G})$, $L(\mathcal{G})$ & set of nodes/links in graph $\mathcal{G}$\\
  \hline
  $|\mathcal{G}|$ & degree of $\mathcal{G}$: $|\mathcal{G}|=|V(\mathcal{G})|$ (number of nodes)\\
  \hline
  $||\mathcal{G}||$ & order of $\mathcal{G}$: $||\mathcal{G}||=|L(\mathcal{G})|$ (number of links)\\
  \hline
  $\mathcal{G} \cup \mathcal{G}^{'} $ & union of graphs: $ \mathcal{G} \cup \mathcal{G}^{'}=(V \cup V^{'}, L \cup L^{'})$\\
  \hline
  $\mathcal{H}$ & interior graph\\
  \hline
  $\mathcal{P}$ & simple path\\
  \hline
  $m_i$ & $m_i\in V(\mathcal{G})$ is the $i$-th monitor in $\mathcal{G}$ \\
  \hline
  $m'_i$ & the $i$-th agent in biconnected component $\mathcal{B}$\\
  \hline
  $\mu_i$ & the $i$-th vantage in triconnected component $\mathcal{T}$\\
  \hline
  $N_{\mathcal{B}}$, $N_{\mathcal{T}}$ & the total number of biconnected/triconnected components in $\mathcal{G}$\\
  \hline
\end{tabular}
\vspace{-.5em}
\end{table}

\section{Auxiliary Algorithms}
To facilitate the presentation of the auxiliary algorithms for DAIL, we first introduce a notion to distinguish between two types of vantages as follows:

\begin{figure}[tb]
\centering
\includegraphics[width=2.7in]{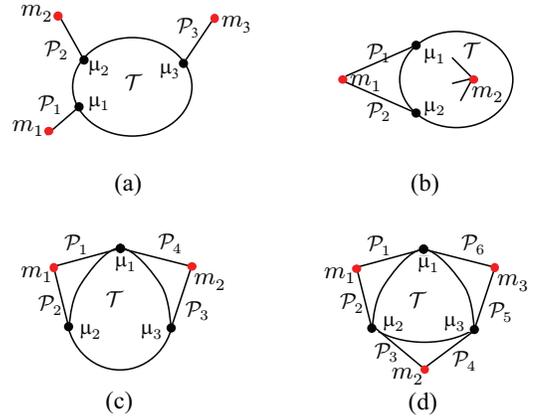}
\vspace{-.5em}
\caption{Four sub-categories of Category 2.\looseness=-1} \label{Fig:AlgSketch}
\vspace{-.3em}
\end{figure}

\begin{definition}\label{def:independent/conjugate vantage}
\emph{\\1)} A vantage $v$ is \emph{conjugate} with another vantage $v'$ (called a \emph{conjugate pair}) if (i) $\{v,\: v'\}$ forms a Type-1-VC, \emph{and} (ii) neither of them is an agent.
\emph{\\2)} A vantage $v$ is \emph{independent} if (i) it is an agent, \emph{or} (ii) it belongs to a Type-1-VC $\{v,\: v'\}$ where $v'$ is a monitor, \emph{or} (iii) it belongs to a Type-$k$-VC with $k\geq 2$.
\end{definition}

The significance of this classification of vantages is that for a given triconnected component, there exists a set of (external) paths (see Lemma~\ref{lemma:vDisjointPaths}) connecting each vantage to an agent such that the path for each independent vantage is vertex disjoint with all the other paths, whereas the paths for vantages in a conjugate pair may lead to the same agent but are disjoint elsewhere (e.g., $\{\mu_1,\mu_2\}$ and $\{\mu_1,\mu_3\}$ in Fig.~\ref{Fig:AlgSketch}(c) are conjugate pairs).
Note that the definition of conjugate pairs only implies that vantages in a conjugate pair must connect to the same agent if their paths are constrained to the neighboring subgraph of the parent biconnected component behind the cut (from the perspective of the triconnected component under consideration). Without this constraint, it is still possible to find disjoint paths for vantages in conjugate pairs, e.g., conjugate vantages $\mu_1$ and $\mu_2$ in Fig.~\ref{Fig:AlgSketch}(c) can connect to agents via disjoint paths $\mathcal{P}_4$ and $\mathcal{P}_2$, respectively.

Based on the type of vantages, we further classify triconnected component of Category 2 into 4 sub-categories:

\vspace{-.5em}
\begin{enumerate}
\item[(i)] \textbf{Category 2.1}: containing only 3 vantages, at least two of which are independent, e.g., $\mathcal{T}$ in Fig.~\ref{Fig:AlgSketch}(a);
\item[(ii)] \textbf{Category 2.2}: containing only 3 vantages, in which only one is independent and the other two form a conjugate pair, e.g., $\mathcal{T}$ in Fig.~\ref{Fig:AlgSketch}(b);
\item[(iii)] \textbf{Category 2.3}: containing only 3 vantages $\{\mu_1,\mu_2,\mu_3\}$, in which none of them is independent, and only $\{\mu_1,\mu_2\}$ and $\{\mu_1,\mu_3\}$ are conjugate pairs, e.g., $\mathcal{T}$ in Fig.~\ref{Fig:AlgSketch}(c);
\item[(iv)] \textbf{Category 2.4}: containing only 3 vantages, in which none of them is independent and each vantage pair forms a conjugate pair, e.g., $\mathcal{T}$ in Fig.~\ref{Fig:AlgSketch}(d);
\end{enumerate}

The strategy of DAIL is to apply Theorem~\ref{theorem:interiorIdentifiability} or \ref{theorem:fullIdentifiability} to each triconnected component together with its associated vantage-to-agent connections. Moreover, the prerequisite of applying Theorem~\ref{theorem:interiorIdentifiability} and \ref{theorem:fullIdentifiability} to the (sub-)graph of interest is that all involved links can be employed for constructing measurement paths. However, for an individual triconnected component, it may involve virtual links, which do not exist in the original graph $\mathcal{G}$. To handle this issue, we prove in Section~\ref{sect:proofs} that for Categories 1, 2.2, 2.4 and 3, even if the associated triconnected component contains virtual links, the conclusions on the sub-graph identifiability in DAIL still hold. However, triangle components of Category 2.1 and 2.3 are special cases requiring separate treatment. Therefore, we introduce auxiliary algorithms, Algorithm~\ref{Alg:triangleC2} and \ref{Alg:triangleC4}. The key question that Algorithm~\ref{Alg:triangleC2} and \ref{Alg:triangleC4} answers is: when virtual links are used in identifying a link, are there real paths in neighboring components to replace these virtual links? Consider a triangle of Category 2.1 (or 2.3) with vantages $\mu_1$, $\mu_2$ and $\mu_3$. To identify link $\mu_2\mu_3$, the path replacement of $\mu_1\mu_2$ or $\mu_1\mu_3$ (if they are virtual) may involve agents, which implies that the condition in Theorem~\ref{theorem:fullIdentifiability} (or Condition \textcircled{\small 2} \normalsize in Theorem~\ref{theorem:interiorIdentifiability}) is not satisfied (see the proof in Section~\ref{sect:proofs}). Therefore, we build the conditions in line~\ref{Tri2:identifiableCondition} of Algorithm~\ref{Alg:triangleC2} (and line~\ref{Tri:conditions} of Algorithm~\ref{Alg:triangleC4}) to ensure the existence of path replacements that do not contain any agents. Specially, if a triconnected component of Category 2.1 satisfies the conditions in line~\ref{Tri2:nonexistenceCondition} (Algorithm~\ref{Alg:triangleC2}), then there is no path replacement of $\mu_1\mu_2$ or $\mu_1\mu_3$, and thus we can at most identify the sum metric of a path and link $\mu_2\mu_3$, but not individual metrics. For a triangle of Category 2.3 with conjugate pairs $\{\mu_1,\mu_2\}$ and $\{\mu_1,\mu_3\}$, nevertheless, any path replacement of $\mu_2\mu_3$ in neighboring components of $\mathcal{T}$ does not involve agents. Thus, this freedom of path selections benefits the identification of $\mu_1\mu_2$ and $\mu_1\mu_3$. In fact, $\mu_1\mu_2$ and $\mu_1\mu_3$ (if any) are always identifiable (line~\ref{Tri:crossLink} of Algorithm~\ref{Alg:triangleC4}). The detailed proof of Algorithms~\ref{Alg:triangleC2} and \ref{Alg:triangleC4} are presented in Section~\ref{sect:proofs}. Since the four sub-categories of Category 2 can cover all Category 2 cases, we have Algorithm~A to identify any triangle component of Category 2.

\begin{algorithm}[tb]
\label{Alg:triangleC2}
\small
\SetKwInOut{Input}{input}\SetKwInOut{Output}{output}
\Input{Triangle component $\mathcal{T}$ of Category 2.1, and the neighboring triconnected components of $\mathcal{T}$}
\Output{Identifiable links in $\mathcal{T}$}
\ForEach{real link $l_i$ in $\mathcal{T}$}
    {
    Suppose $l_i$ is incident to $v_1$ and $v_2$ and the third node in $\mathcal{T}$ is $v_3$. Let $S_1$ ($S_2$) be the set of immediate neighboring triconnected components connected to $\mathcal{T}$ via $\{v_1,v_3\}$ ($\{v_2,v_3\}$)\;
    Let $\mathcal{T}^*_1$ ($\mathcal{T}^*_2$) be a neighboring triconnected component of $\mathcal{T}$ with the following properties: (i) $v_1\in \mathcal{T}^*_1$ ($v_2\in \mathcal{T}^*_2$); (ii) $v_2\notin \mathcal{T}^*_1$ ($v_1\notin \mathcal{T}^*_2$); (iii) $\mathcal{T}^*_1$ ($\mathcal{T}^*_2$) contains at least two real links which are incident to $v_1$ ($v_2$); (iv) $\mathcal{T}^*_1$ ($\mathcal{T}^*_2$) contains 3 or more vantages\;\label{Tri2:TstarProperties}
    \ForEach{$j$ ($j=1,2$)}
        {
        \If{$v_jv_3$ is virtual \textbf{AND} $v_j$ is not an agent in the parent biconnected component of $\mathcal{T}$ \textbf{AND} $\{v_1,v_2\}$ is a Type-0-VC w.r.t. $\mathcal{T}$ if $\{v_1,v_2\}$ is a 2-vertex cut \textbf{AND} $|S_j|=1$ \textbf{AND} $\mathcal{T}^*_j$ satisfying the properties in line~\ref{Tri2:TstarProperties} does not exist\label{Tri2:nonexistenceCondition}}
        {
           $l_i$ is unidentifiable; break\; \label{Tri2:Unidentifiable}
        }
        }
        \If{$l_i$ is not marked as unidentifiable by line~\ref{Tri2:Unidentifiable}\label{Tri2:identifiableCondition}}
        {
        $l_i$ is identifiable\;
        }
    }
\caption{Determination of All Identifiable Links in Triangles of Category 2.1}
\vspace{-.25em}
\end{algorithm}
\normalsize

\begin{algorithm}[tb]
\label{Alg:triangleC4}
\small
\SetKwInOut{Input}{input}\SetKwInOut{Output}{output}
\Input{Triangle component $\mathcal{T}$ of Category 2.3 with two conjugate pairs $\{\mu_1,\mu_2\}$ and $\{\mu_1,\mu_3\}$, and a set $S_1$ ($S_2$) of its immediately neighboring triconnected components connected to $\mathcal{T}$ via $\{\mu_1,\mu_2\}$ ($\{\mu_1,\mu_3\}$)}
\Output{Identifiable links in $\mathcal{T}$}
$\mu_1\mu_2$ ($\mu_1\mu_3$) is identifiable if it is a real link\;\label{Tri:crossLink}
\If{(link $\mu_1\mu_2$ is real \textbf{OR} $|S_1|\geq 2$ \textbf{OR} one component in $S_1$ is 3-vertex-connected) \textbf{AND} (link $\mu_1\mu_3$ is real \textbf{OR} $|S_2|\geq 2$ \textbf{OR} one component in $S_2$ is 3-vertex-connected)\label{Tri:conditions}}
{$\mu_2\mu_3$ is identifiable if it is a real link\;}
\caption{Determination of All Identifiable Links in Triangles of Category 2.3}
\vspace{-.25em}
\end{algorithm}
\normalsize

\newcounter{hdps}
\stepcounter{hdps}

\begin{algorithm}[tb]
\renewcommand\thealgocf{\Alph{hdps}}
\label{Alg:OVERALLtriangle}
\small
\SetKwInOut{Input}{input}\SetKwInOut{Output}{output}
\Input{Triangle component $\mathcal{T}$ of Category 2 and its neighboring triconnected components}
\Output{Identifiable links in $\mathcal{T}$}
\uIf{$\mathcal{T}$ is of Category 2.1}
{the identifiability of $\mathcal{T}$ is determined by Algorithm~\ref{Alg:triangleC2}\;}
\uElseIf{$\mathcal{T}$ is of Category 2.2}
{all links except for the ones incident to the independent vantage in $\mathcal{T}$ are identifiable\;}
\uElseIf{$\mathcal{T}$ is of Category 2.3}
{the identifiability of $\mathcal{T}$ is determined by Algorithm~\ref{Alg:triangleC4}\;}
\Else(\texttt{// $\mathcal{T}$ must be of Category 2.4})
{all links in $\mathcal{T}$ are identifiable\;}
\caption{Determination of All Identifiable Links in Triangles of Category 2}
\vspace{-.25em}
\end{algorithm}
\normalsize

\stepcounter{hdps}

\begin{algorithm}[tb]
\renewcommand\thealgocf{\Alph{hdps}}
\label{Alg:C6}
\small
\SetKwInOut{Input}{input}\SetKwInOut{Output}{output}
\Input{Triconnected component $\mathcal{T}$ of Category 3 with direct link $\mu_1\mu_2$, and the neighboring biconnected components connected to $\mathcal{T}$ through $\mu_1$ or $\mu_2$}
\Output{Identifiability of link $\mu_1\mu_2$}
\uIf {there exist neighboring triconnected components connecting to $\mathcal{T}$ by 2-vertex cut $\{\mu_1,\mu_2\}$ \textbf{AND} $\mu_1$ and $\mu_2$ are not the (only two) vantages in one of these neighboring components\label{C6:AlreadyIdentifiable}}
    {
    $\mu_1\mu_2$ is determined in neighboring triconnected components\;
    }
\Else
    {
    Let $S^{\mathcal{B}}_1$ ($S^{\mathcal{B}}_2$) be the set of biconnected components containing $\mu_1$ ($\mu_2$) as the only common node with $\mathcal{T}$, and $n^{\mathcal{B}}_1$ ($n^{\mathcal{B}}_2$) be the total number of agents (excluding $\mu_1$ and $\mu_2$) w.r.t. each biconnected component in $S^{\mathcal{B}}_1$ ($S^{\mathcal{B}}_2$)\;
    \uIf{($\mu_1$ is a real monitor \textbf{OR} $n^{\mathcal{B}}_1\geq 2$) \textbf{AND} ($\mu_2$ is a real monitor \textbf{OR} $n^{\mathcal{B}}_2\geq 2$)\label{C6:IdentifiableCondition}}
    {
    $\mu_1\mu_2$ is identifiable\;
    }
    \Else
    {$\mu_1\mu_2$ is unidentifiable\;}
    }
\caption{Determination of Direct Links in Components of Category 3}
\vspace{-.25em}
\end{algorithm}
\normalsize

\vspace{-.5em}
Lemma~\ref{lemma:effectiveMonitor} states that the identification of direct link $m'_1m'_2$ cannot be immediately determined in a biconnected component with only 2 agents $m'_1$ and $m'_2$. Similarly, the direct link $\mu_1\mu_2$ in a triconnected component $\mathcal{T}$ with only 2 vantages ($\mu_1$ and $\mu_2$) cannot be identified by purely network internal structure of $\mathcal{T}$, i.e., external connections between $\mathcal{T}$ and other neighboring components have to be employed for identifying $\mu_1\mu_2$ in $\mathcal{T}$. To this end, we develop Algorithm~B, the correctness of which is guaranteed by Lemma.~\ref{lemma:Alg5} in Section~\ref{sect:theorems}.


\section{Theorems}
\label{sect:theorems}

\begin{theorem}
\label{theorem:unidentifiable}
\cite{Ma13IMC} If only two monitors $m_1$ and $m_2$ are used, then none of the exterior links, except for link $m_1m_2$, is identifiable.
\end{theorem}

\begin{theorem}
\label{theorem:interiorIdentifiability}
\cite{Ma13IMC} If only two monitors $m_1$ and $m_2$ are used, the interior graph $\mathcal{H}$ (with $|L(\mathcal{H})|>1$) of $\mathcal{G}$ is connected and link $m_1m_2$ does not exist, then the necessary and sufficient conditions for identifying \emph{all} link metrics in $\mathcal{H}$ are:

\vspace{-.5em}
\begin{description}
  \item[\textcircled{\small 1}]\normalsize The remaining graph after deleting any interior link in $\mathcal{G}$ is 2-edge-connected;
  \item[\textcircled{\small 2}]\normalsize The augmented graph after adding link $m_1m_2$ to $\mathcal{G}$ is 3-vertex-connected.
\end{description}
\end{theorem}

\begin{theorem}
\label{theorem:fullIdentifiability}
\cite{Ma13IMC} Using $\kappa$ ($\kappa\geq 3$) monitors, $\mathcal{G}$ is completely identifiable \emph{if and only if} the associated {extended graph $\mathcal{G}_{ex}$} is 3-vertex-connected.
\end{theorem}

\begin{lemma}
\label{lemma:effectiveMonitor}
Let $\mathcal{B}$ be a biconnected component with agents $m'_1,\ldots,m'_{\kappa}$. The identifiability of links in $\mathcal{B}$ does not depend on whether $m'_1,\ldots,m'_{\kappa}$ are monitors or not, except for link $m'_1m'_2$ (if it exists) when $\kappa=2$.
\end{lemma}

\vspace{-.5em}
\noindent\textbf{Claim 1.} A triconnected component $\mathcal{T}$ may contain multiple virtual links. For each involved virtual link whose end-points $\{v_1,v_2\}$ (the end-points of a virtual link must form a vertex cut) form a Type-0-VC wrt $\mathcal{T}$, there exists a simple path $\mathcal{P}_r$ with the same end-points in a neighboring biconnected component $\mathcal{B}_{\mathcal{T}}$ which connects to $\mathcal{T}$ via $\{v_1,v_2\}$. $\mathcal{P}_r$ can be used to replace the associated virtual link in $\mathcal{T}$ if this virtual link is chosen to construct measurement paths for identifying real links in $\mathcal{T}$. This replacement operation does not affect all existing path construction policies or the identifiability of real links in $\mathcal{T}$. Such $\mathcal{P}_r$ also exists if $\{v_1,v_2\}$ forms a Type-$k$-VC ($k\geq 1$), but no agents (excluding $v_1$ and $v_2$) in $\mathcal{B}_{\mathcal{T}}$ are used for identifying $\mathcal{T}$.

\begin{lemma}
\label{lemma:vDisjointPaths}
For all vantages $\{\mu_i\}$ in a triconnected component $\mathcal{T}$ of $\mathcal{G}$, there exists an external $\mu_i$-to-agent path $\mathcal{P}$ ($\mathcal{P}$ is a degenerated single node if $\mu_i$ is an agent) which is internally vertex disjoint with all other $\mu_j$-to-agent paths ($j\neq i$).
\end{lemma}

\begin{lemma}\label{lemma:complete classification}
Any triconnected component within a biconnected component with two or more agents belongs to one of \textbf{Categories 1--3}.
\end{lemma}

\begin{lemma}
\label{lemma:Alg3}
Algorithm~\ref{Alg:triangleC2}, Determination of All Identifiable Links in Triangles of Category 2.1, can determine all identifiable links in a triangle of Category 2.1.
\end{lemma}

\begin{lemma}
\label{lemma:Alg4}
Algorithm~\ref{Alg:triangleC4}, Determination of All Identifiable Links in Triangles of Category 2.3, can determine all identifiable links in a triangle of Category 2.3.
\end{lemma}

\begin{lemma}
\label{lemma:Alg5}
Algorithm~B, Determination of Direct Links in Components of Category 3, can determine the identifiability of the direct link in a triconnected component of Category 3.
\end{lemma}

\begin{theorem}
\label{theorem:uniqueTriNum}
For the triconnected component decomposition of graph $\mathcal{G}$, (i) the 3-vertex-connected and single-link components are unique, while triangle components may not be unique; and (ii) the total number of triconnected components $\mathcal{N}_{\mathcal{T}}$ in $\mathcal{G}$, however, is a constant.
\end{theorem}

\begin{theorem}
\label{theorem:FLIcorrectness}
Algorithm DAIL \cite{MaInfocom14sub}, Determination of All Identifiable Links, can determine all identifiable and unidentifiable links in $\mathcal{G}$ with given monitor placement.
\end{theorem}

\begin{theorem}
\label{theorem:candidates}
The candidate set $\mathcal{S}$ selected by Algorithm \cite{MaInfocom14sub} Candidate Monitor Selection always contains an optimal monitor placement as a subset.
\end{theorem}

Let $\mathcal{M}$ denote a given minimum set of monitors for complete network identification as selected by MMP\footnote{This set may not be unique, but the number of nodes in $\mathcal{M}$, i.e., $|\mathcal{M}|$, is a constant and our argument in Theorem~\ref{theorem:optimal2VC} holds for any such $\mathcal{M}$.}. Let $\mathcal{O}^*_{\kappa}$ denote an optimal $\kappa$-monitor placement in $\mathcal{G}$ ($\mathcal{O}^*_{\kappa}$ may not be unique). We have the following theorem.

\begin{theorem}
\label{theorem:optimal2VC}
If $\mathcal{G}$ is 2-vertex-connected and $3\leq \kappa < |\mathcal{M}|$, then (i) $\exists \mathcal{O}^*_\kappa$ such that $\mathcal{O}^*_{\kappa}\subseteq \mathcal{M}$, and (ii) for any given $\mathcal{O}^*_\kappa$ with $\mathcal{O}^*_{\kappa}\subseteq \mathcal{M}$, $\exists \mathcal{O}^*_{\kappa+1}$ such that $\mathcal{O}^*_{\kappa+1}=\mathcal{O}^*_{\kappa}\cup \{v_{m}\}$, where $v_{m}={\operatorname*{arg\max}}_{v}N(\mathcal{O}^*_{\kappa}\cup \{v\})$ over $v \in \mathcal{M}\setminus \mathcal{O}^*_{\kappa}$.
\end{theorem}

\section{Proofs}
\label{sect:proofs}

\subsection{Proof of Lemma~\ref{lemma:effectiveMonitor}}
\subsubsection{Consider the case with $\kappa\geq 3$}
\begin{figure}[tb]
\centering
\includegraphics[width=2.6in]{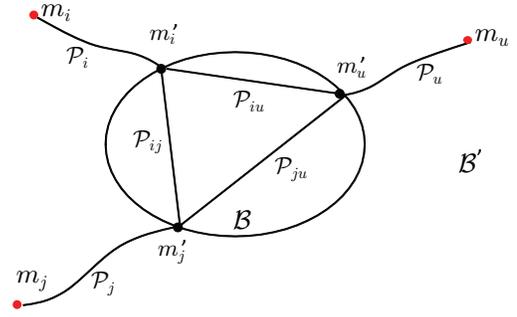}
\caption{Biconnected component $\mathcal{B}$ with 3 or more agents.}\label{fig:EffectiveMonitorK3}
\vspace{-.5em}
\end{figure}
Suppose biconnected component $\mathcal{B}$ in Fig.~\ref{fig:EffectiveMonitorK3} contains 3 or more agents. For any simple path $\mathcal{P}_{ij}$ connecting two agents $m'_i$ and $m'_j$ within $\mathcal{B}$, i.e, $V(\mathcal{P}_{ij})\in V(\mathcal{B})$, it suffices to show that path metric $W_{\mathcal{P}_{ij}}$ can be calculated by path measurements between real monitors. Employing nodes within $\mathcal{B}$, there exist two internally vertex disjoint paths $\mathcal{P}_{iu}$ and $\mathcal{P}_{ju}$ connecting $m'_i$ and $m'_j$ to another agent $m'_u$ (the total agent number $\geq$ 3) with $V(\mathcal{P}_{ij}\cap\mathcal{P}_{iu})=\{m'_i\}$ and $V(\mathcal{P}_{ij}\cap\mathcal{P}_{ju})=\{m'_j\}$. This is because if any two of these three paths must have a common node (except for the common end-point), then this common node is a cut-vertex, contradicting the property of $\mathcal{B}$ being 2-vertex-connected. In addition, there exist three vertex disjoint paths $\mathcal{P}_i$, $\mathcal{P}_j$ and $\mathcal{P}_u$, each connecting an agent and a real monitor (see Fig.~\ref{fig:EffectiveMonitorK3}). Abstracting $\mathcal{P}_i$, $\mathcal{P}_j$, $\mathcal{P}_u$, $\mathcal{P}_{ij}$, $\mathcal{P}_{iu}$ and $\mathcal{P}_{ju}$ as single links, the augmented graph $\mathcal{B}'$ containing these six links and six nodes ($m'_i$, $m'_j$, $m'_u$, $m_i$, $m_j$, and $m_u$) satisfies the condition in Theorem~III.3, and thus $\mathcal{B}'$ is fully identifiable. Therefore, path metric of $W_{\mathcal{P}_{ij}}$ can be calculated by path measurements between real monitors when $\kappa\geq 3$.

\subsubsection{Consider the case with $\kappa=2$}

\begin{figure}[tb]
\centering
\includegraphics[width=2.2in]{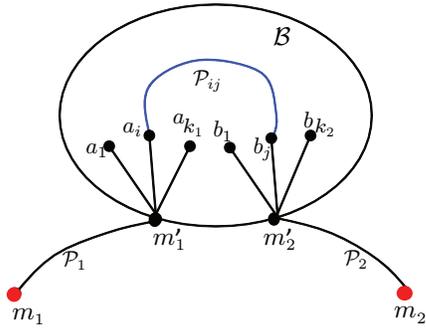}
\vspace{-.5em}
\caption{Agents $m'_1$ and $m'_2$ wrt biconnected component $\mathcal{B}$.}\label{fig:EffectiveMonitorProof}
\vspace{-.5em}
\end{figure}
Let $m'_1$ and $m'_2$ be the two agents of biconnected component $\mathcal{B}$ in Fig.~\ref{fig:EffectiveMonitorProof} and $m'_1$ ($m'_2$) connects to the real monitor $m_1$ ($m_2$) by path\footnote{$\mathcal{P}_1$ and $\mathcal{P}_2$ may not be unique.} $\mathcal{P}_1$ ($\mathcal{P}_2$), i.e., none of $m'_1$ and $m'_2$ are real monitors. In Fig.~\ref{fig:EffectiveMonitorProof}, it is impossible that $\mathcal{P}_1$ and $\mathcal{P}_2$ must have a common node; since otherwise $m'_1$ and $m'_2$ are not cut-vertices, contradicting the processing of localizing agents for a biconnected component (see Algorithm DAIL in \cite{MaInfocom14sub}). To identify link metrics in $\mathcal{B}$, all measurement paths involving links in $\mathcal{B}$ are of the following form
\begin{equation}\label{eq:effectiveMonitorEqForm}
    W_{\mathcal{P}_1}+W_{m'_1a_i}+W_{\mathcal{P}_{ij}}+W_{b_jm'_2}+W_{\mathcal{P}_2}=c'_{ij},
\end{equation}
assuming $\mathcal{P}_1$ ($\mathcal{P}_2$) is always selected to connect $m_1$ and $m'_1$ ($m_2$ and $m'_2$). We know that if $m'_1$ and $m'_2$ are real monitors, then each measurement (except direct link $m'_1m'_2$) path is of form
\begin{equation}\label{eq:MonitorEqForm}
    W_{m'_1a_i}+W_{\mathcal{P}_{ij}}+W_{b_jm'_2}=c_{ij}.
\end{equation}
Therefore, compared with (\ref{eq:MonitorEqForm}), (\ref{eq:effectiveMonitorEqForm}) is equivalent to abstracting each of ${\mathcal{P}_1}+{m'_1a_i}$ and ${b_jm'_2}+{\mathcal{P}_2}$ as a single link. By Theorem~III.1, we know that none of the  exterior links are identifiable. Thus, the link metrics of exterior links do not affect the identification of interior links. Therefore, $\mathcal{B}$ can be visualized as a network with two monitors $m'_1$ and $m'_2$ but each exterior link in $\{\{m'_1a_i\},\{m'_2b_j\}\}$ has an added weight from $W_{\mathcal{P}_1}$ or $W_{\mathcal{P}_2}$. The above argument also holds when $m_1$ ($m_2$) chooses another path, say $\mathcal{P}'_1$ ($\mathcal{P}'_2$), to connect to $m'_1$ ($m'_2$), then it simply implies that different exterior links in $\{\{m'_1a_i\},\{m'_2b_j\}\}$ in $\mathcal{B}$ may have different added path weights when regarding $m'_1$ and $m'_2$ as two monitors. Moreover, the above conclusion also applies to the case that one of $m'_1$ and $m'_2$ is a real monitor. Therefore, the identifiability of all links except for the direct link $l_d=m'_1m'_2$ (if any) remains the same regardless of whether $m'_1$, $m'_2$ are monitors or not.\looseness=-1
\hfill$\blacksquare$

\subsection{Proof of Claim 1}
\begin{figure}[tb]
\centering
\includegraphics[width=1.7in]{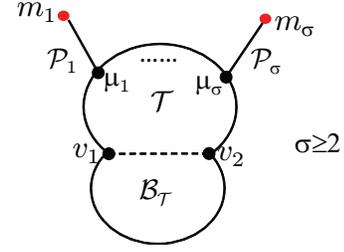}
\vspace{-.5em}
\caption{Virtual link replacement.}\label{fig:Claim1}
\vspace{-.5em}
\end{figure}

Fig.~\ref{fig:Claim1} illustrates a triconnected component $\mathcal{T}$ with $\sigma$ vantages ($\sigma\geq 2$), where some of these $\sigma$ vantages may form conjugate pairs. For vertex cut $\{v_1,v_2\}$ (which is a Type-0-VC), there exists a simple path $\mathcal{P}_r$ connecting $v_1$ and $v_2$ in the neighboring biconnected component $\mathcal{B}_\mathcal{T}$ within the parent biconnected component of $\mathcal{T}$ as $\mathcal{B}_\mathcal{T}$ contains at least 3 nodes. We know that $\mathcal{B}_\mathcal{T}$ connects to $\mathcal{T}$ via only $\{v_1,v_2\}$; therefore, $\mathcal{P}_1,\ldots,\mathcal{P}_{\sigma}$ do not have common nodes with $\mathcal{B}_\mathcal{T}$ except that $\mu_i$ ($i=1,\ldots,\sigma$) may equal $v_1$ ($v_2$). Hence, for virtual link $v_1v_2$, if it is used for identifying real links in $\mathcal{T}$ according to Theorem III.2 or III.3, then it can be replaced by $\mathcal{P}_r$ which is a simple path and can be abstracted as a real link in $\mathcal{T}$.

For the case that $\{v_1,v_2\}$ forms a Type-$k$-VC ($k\geq 1$) and no agents (excluding $v_1$ and $v_2$) in $\mathcal{B}_{\mathcal{T}}$ are used for identifying $\mathcal{T}$, we can ignore these agents in $\mathcal{B}_{\mathcal{T}}$. Accordingly, this Type-$k$-VC is converted to a Type-0-VC. Therefore, the above argument applies to this special case, suggesting the existence of replacement path $\mathcal{P}_r$.
\hfill$\blacksquare$

\subsection{Proof of Lemma~\ref{lemma:vDisjointPaths}}

\emph{1)} If vantage $\mu_i$ is an agent, then $\mu_i$-to-agent path is itself $\mu_i$ (i.e., a path containing only one node). This is a trivial case that $\mu_i$-to-agent path is internally vertex disjoint with all other vantage-to-agent paths.

\emph{2)} For triconnected component $\mathcal{T}$ in biconnected component $\mathcal{B}$, if vantage $\mu_i$ is not an agent and $\mu_i$ is in a Type-1-VC, then let $\mu'_i$ be the other node in this Type-1-VC, $\mathcal{P}$ the $\mu_i$-to-agent path, and $\mathcal{P}'$ the $\mu'_i$-to-agent path. If $\mu_i$ and $\mu'_i$ form a conjugate pair, then $\mathcal{P}$ and $\mathcal{P}'$ may have a common terminating-point (i.e., the agent). In this case, in addition to the common terminating-point $m'$, if $\mathcal{P}$ and $\mathcal{P}'$ must have another common node $w$ (an internal vertex in $\mathcal{P}$ and $\mathcal{P}'$), then it implies that $\mu_i$ and $\mu'_i$ cannot connect to $m'$ if $w$ is deleted. $w$ therefore is a cut-vertex in $\mathcal{B}$, contradicting the assumption that $\mathcal{B}$ is a biconnected component. If $\mu_i$ and $\mu'_i$ do not form a conjugate pair, then $\mu'_i$ is an agent. In this case,
$\mu'_i$-to-agent path is $\mu'_i$, which is internally vertex disjoint with $\mathcal{P}$.

\emph{3)} For triconnected component $\mathcal{T}$ in biconnected component $\mathcal{B}$, if vantage $\mu_i$ is not an agent and $\mu_i$ is in a Type-$k$-VC ($k\geq 2$), then we can always construct $\mu_i$-to-$m'$ path $\mathcal{P}$ and $\mu'_i$-to-$m'$ path $\mathcal{P}'$ ($\mu'_i$ is another node in this Type-$k$-VC, suppose $\mu'_i$ is not an agent) to make sure that $\mathcal{P}$ and $\mathcal{P}'$ have the same terminating-point $m'$. Then the same argument in 2) applies to this case.

\emph{4)} In 2) and 3), we only consider constructing $\mathcal{P}$ and $\mathcal{P}'$ in the same neighboring biconnected component $\mathcal{B}_N$ of $\mathcal{T}$. For the vantage-to-agent paths in other neighboring biconnected component of $\mathcal{T}$, these paths do not use any nodes in $\mathcal{B}_N$ (except for the possible case that the starting-point, i.e., the vantage, is in $\mathcal{B}_N$). Thus, these vantage-to-agent paths are still internally vertex disjoint with $\mathcal{P}$.

Therefore, for vantages in the same Type-$k$-VC, their vantage-to-agent paths at most have one common terminating-point (i.e., the agent). While for vantages in different Type-$k$-VCs, their vantage-to-agent paths at most have a common starting-point (i.e., the vantage). Consequently, For all vantages $\{\mu_i\}$ in $\mathcal{T}$, there exists an external $\mu_i$-to-agent path $\mathcal{P}$ which is internally vertex disjoint with all other $\mu_j$-to-agent paths ($j\neq i$).\hfill$\blacksquare$

\subsection{Proof of Lemma~\ref{lemma:complete classification}}
It is easy to see that \textbf{Categories 1--3} are mutually exclusive and cover all possibilities except for the case containing 0 or 1 vantage.
If $\mathcal{T}$ only has 0 or 1 vantage, then $\mathcal{B}$ can only have 0 or 1 agent located in $\mathcal{T}$, as any other agent outside $\mathcal{T}$ will imply at least two vantages in $\mathcal{T}$ (nodes in the 2-vertex cut separating $\mathcal{T}$ from the agent must be vantages).
Therefore, any triconnected component with the parent biconnected component containing 2 or more agents falls into one of Categories 1--3.
\hfill$\blacksquare$

\subsection{Proof of Lemma~\ref{lemma:Alg3}}
For real link $l_i$ in triconnected component $\mathcal{T}$, let $v_1$ and $v_2$ be the end-points of $l_i$, $v_3$ the third node in $\mathcal{T}$, $S_1$ ($S_2$) the set of immediately neighboring triconnected components connected to $\mathcal{T}$ via $\{v_1,v_3\}$ ($\{v_2,v_3\}$), and $S^*_1$ ($S^*_2$) the set of immediately neighboring biconnected components connected to $\mathcal{T}$ via $\{v_1,v_3\}$ ($\{v_2,v_3\}$) within the same parent biconnected component. Note all links in $S^*_1$ ($S^*_2$) except for $v_1v_3$ ($v_2v_3$) are real links. It suffices to show how the identifiability of $l_i$ is determined by Algorithm~\ref{Alg:triangleC2}. Let $\mathcal{T}^*_1$ ($\mathcal{T}^*_2$) be a neighboring triconnected component of $\mathcal{T}$ with the following properties: (i) $v_1\in \mathcal{T}^*_1$ ($v_2\in \mathcal{T}^*_2$); (ii) $v_2\notin \mathcal{T}^*_1$ ($v_1\notin \mathcal{T}^*_2$); (iii) $\mathcal{T}^*_1$ ($\mathcal{T}^*_2$) contains at least two real links which are incident to $v_1$ ($v_2$); (iv) $\mathcal{T}^*_1$ ($\mathcal{T}^*_2$) contains 3 or more vantages.

We first consider the condition (called Condition A in the sequel) that: (a) $v_1v_3$ is virtual \textbf{AND} (b) $v_1$ is not an agent in the parent biconnected component of $\mathcal{T}$ \textbf{AND} (c) $\{v_1,v_2\}$ is a Type-0-VC w.r.t. $\mathcal{T}$ if $\{v_1,v_2\}$ is a 2-vertex cut \textbf{AND} (d) $|S_1|=1$ \textbf{AND} (e) $\mathcal{T}^*_1$ satisfying the properties (i)--(iv) does not exist. We show that if $\mathcal{T}$ does not satisfy any of the five conditions (a)--(e) in Condition A, then there exists a path replacement for $v_1v_3$ (denoted by $\mathcal{P}(v_1,v_r)$), one $v_1$-to-agent path, and one $v_r$-to-agent path, with all these three paths being internally vertex disjoint.

\begin{figure}[tb]
\centering
\includegraphics[width=1.9in]{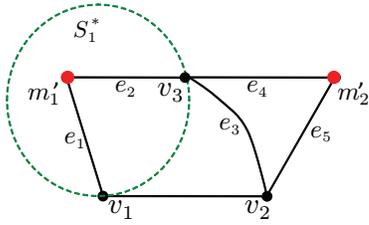}
\vspace{-.5em}
\caption{Triangle identification ($v_1$ is an agent).}\label{Fig:v1v2agent}
\vspace{-.5em}
\end{figure}

\begin{enumerate}[1)]
  \item If $v_1v_3$ is a real link, then it is obvious that there is no need to find a path replacement for $v_1v_3$. Moreover, according to the definition of Category 2.1, there exist two vantage-to-agent (i.e., $v_1$-to-agent and $v_3$-to-agent) disjoint paths;
  \item If $v_1$ is an agent in the parent biconnected component of $\mathcal{T}$, then $v_1$ itself is a degenerated $v_1$-to-agent path (containing a single node). In this case, if $S^*_1$ contains one agent (excluding $v_1$), say $m'_1$ (as illustrated in Fig.~\ref{Fig:v1v2agent}), then there exist internally vertex disjoint paths $m'_1e_1v_1$ and $m'_1e_2v_3$ since each component in $S^*_1$ is 2-vertex-connected. Hence, $m'_1e_1v_1$ can be used as the path replacement for $v_1v_3$. Note that $m'_1e_2v_3$ might be merged with another path for identifying $v_1v_2$, which will be clear in later discussions. If $S^*_1$ does not contain any agents except for $v_1$, then there exists path replacement connecting $v_1$ and $v_3$ within $S^*_1$. Furthermore, for this replacement path, the end-point $v_3$ has a vantage-to-agent path in $S^*_2$ or $v_3$ itself is an agent according to the definition of Category 2.1.

\begin{figure}[tb]
\centering
\includegraphics[width=1.7in]{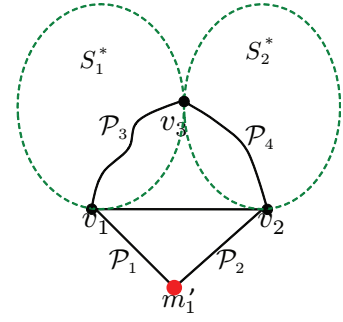}
\vspace{-.5em}
\caption{Triangle identification ($v_1v_2$ is a cross-link).}\label{Fig:v1v2crossLink}
\vspace{-.5em}
\end{figure}

  \item If $\{v_1,v_2\}$ is a Type-$k$-VC ($k\geq 1$) w.r.t. $\mathcal{T}$ when $\{v_1,v_2\}$ is a 2-vertex cut, then, as illustrated in Fig.~\ref{Fig:v1v2crossLink}, $v_1v_2$ is a Cross-link \cite{Ma13IMC}, which is identifiable using paths $\mathcal{P}_1,\ldots,\mathcal{P}_4$ (see \cite{Ma13IMC}).
  \item If $|S_1|>1$, then it implies $|S^*_1|>1$. Thus, the path replacement for $v_1v_3$ can be chosen from one component in $S^*_1$ and there exist $v_1$-to-agent ($v_3$-to-agent) path in another component of $S^*_1$.

\begin{figure}[tb]
\centering
\includegraphics[width=2.6in]{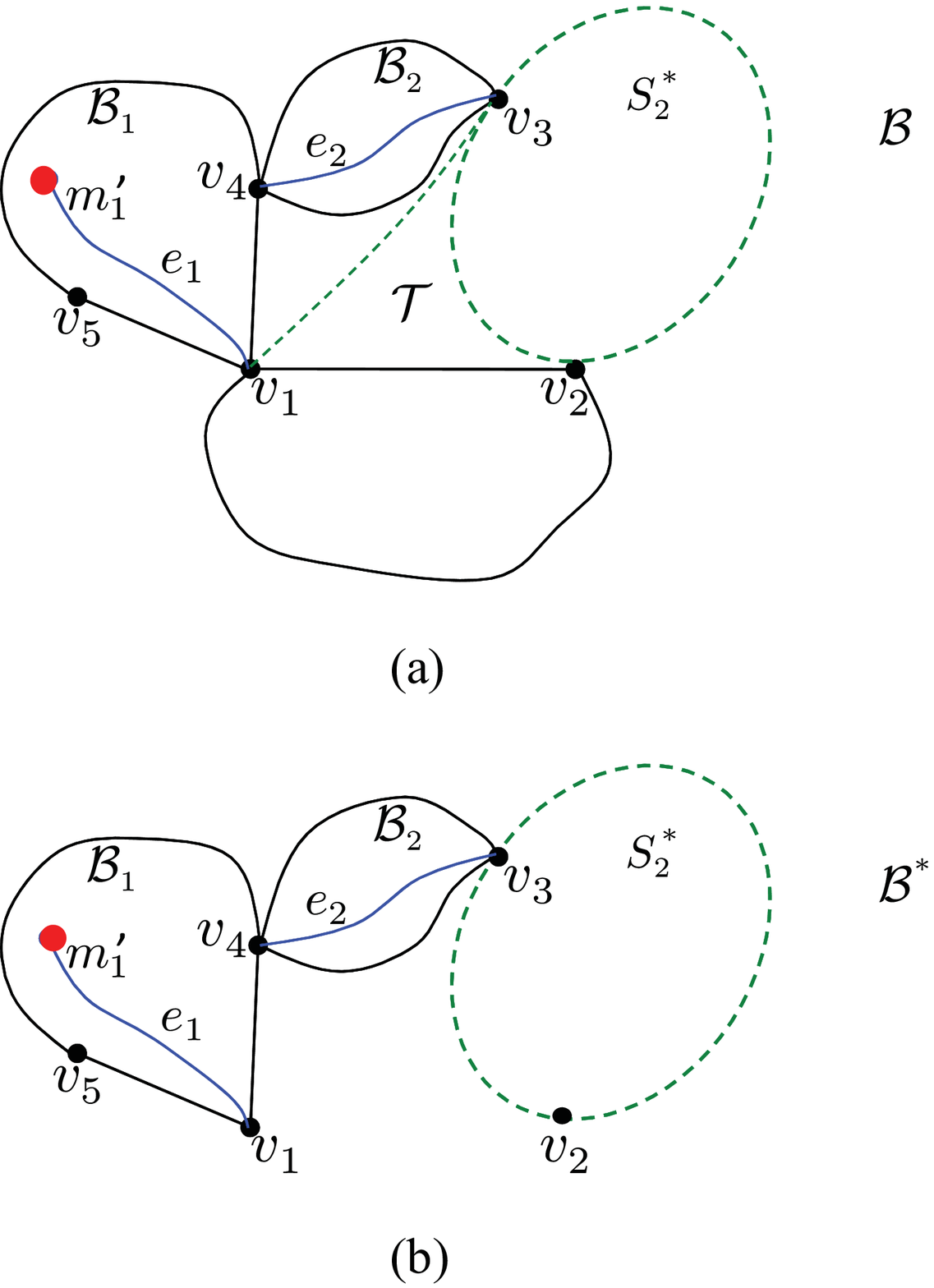}
\vspace{-.5em}
\caption{Triangle identification (there exists $\mathcal{T}^*$ in $\mathcal{B}_1$).}\label{Fig:quasiBC}
\vspace{-.5em}
\end{figure}

  \item Suppose there exists $\mathcal{T}^*_1$ satisfying the properties in (i)--(iv). This is the most complicated condition in Condition A. Let $\mathcal{B}$ denote the parent biconnected component of $\mathcal{T}$. If we remove the neighboring biconnected component (except for nodes $v_1$ and $v_2$) connecting to $\mathcal{T}$ via $v_1v_2$, then, as illustrated in Fig.\ref{Fig:quasiBC}, $v_3$ is a cut-vertex in the remaining graph $\mathcal{B}^*$. In $\mathcal{B}^*$, for the connected component containing $v_1$ and $v_3$, there exists a subgraph $\mathcal{B}_1$ which is 2-vertex-connected or a bond\footnote{A bond is a graph with only two nodes and one link connecting these two nodes.}, as illustrated in Fig.\ref{Fig:quasiBC}. Note that in Fig.\ref{Fig:quasiBC}, all links in $\mathcal{B}_1$ and $\mathcal{B}_2$ are real links since they are 2-vertex-connected (or bonds), and $\mathcal{B}_1$ contains $\mathcal{B}_2$ when $v_3$ is the only cut-vertex in $\mathcal{B}^*$. Suppose (a)--(d) in Condition A are satisfied. If there exist real links $v_1v_4$ and $v_1v_5$ in $\mathcal{B}_1$, then $\mathcal{T}^*_1$ must be a subgraph in $\mathcal{B}_1$. Since $v_2$ and $v_3$ are vantages of $\mathcal{T}$, $v_1$ and $v_4$ must be the two vantages of $\mathcal{T}^*_1$. If $\mathcal{T}^*_1$ contains a third vantage, then it implies that there exists an agent $m'_1$ (with $m'_1\neq v_1$, $m'_1\neq v_4$) in $\mathcal{B}_1$. Hence, as $\mathcal{B}_1$ is 2-vertex-connected, there exist internally vertex disjoint paths $v_1e_1m'_1$ and $v_1v_4e_2v_3$. Therefore, $v_1v_4e_2v_3$ is chosen as the path replacement for $v_1v_3$. Moreover, there exist $v_3$-to-agent path (guaranteed by Condition B discussed as follows) in $S^*_2$ or $v_3$ itself is an agent in $\mathcal{B}$.
\end{enumerate}
Let Condition B be (a) $v_2v_3$ is virtual \textbf{AND} (b) $v_2$ is not an agent in the parent biconnected component of $\mathcal{T}$ \textbf{AND} (c) $\{v_1,v_2\}$ is a Type-0-VC w.r.t. $\mathcal{T}$ if $\{v_1,v_2\}$ is a 2-vertex cut \textbf{AND} (d) $|S_2|=1$ \textbf{AND} (e) $\mathcal{T}^*_2$ satisfying the properties in (i)--(iv) does not exist. The above argument in 1)--5) also applies to the path replacement for $v_2v_3$. If neither Condition A nor Condition B is satisfied, then the two path replacements for $v_1v_3$ and $v_2v_3$ can be abstracted as two single links, thus forming a triangle containing link $v_1v_2$. Moreover, there exist vertex disjoint vertex-to-agent paths for the three vertices in this constructed triangle. Therefore, $v_1v_2$ is identifiable since the resided triangle with three vertex-to-agent paths (abstracting paths as single links) satisfies the condition in Theorem III.3.

\emph{Remark}: When neither Condition A nor Condition B is satisfied, there are two other possible scenarios. (i) As shown in Fig.~\ref{Fig:v1v2agent}, if the replacement path for $v_2v_3$ is $v_2e_3v_3$, then $m'_1e_2v_3+v_3e_3v_2$ can be abstracted as a single link in a triangle; (ii) In the case that $v_1$ and $v_2$ are both agents, if the replacement path for $v_2v_3$ is $m'_2e_5v_2$ or $v_3$ itself is an agent, then a quadrangle (i.e., $(\{m'_1,v_1,v_2,m'_2\},\{m'_1e_1v_1,v_1v_2,v_2e_5m'_2,m'_1v_3m'_2\})$ or $(\{m'_1,v_1,v_2,v_3\},\{m'_1e_1v_1,v_1v_2,v_2e_3v_3,m'_1e_2v_3\}$, where paths are abstracted as single links) is formed. Nevertheless, this quadrangle with its vertex-to-agent paths (abstracting paths as single links) also satisfies the condition in Theorem III.3.

If (a)--(e) in Condition A are satisfied, then it implies that $|S^*_1|=1$ and there is no agent (except for $v_4$) in $\mathcal{B}_1$ (see Fig.~\ref{Fig:quasiBC}). Therefore, we can at most identify the sum metric of $v_1v_2$ and a path within $\mathcal{B}_1$, but not the individual link metric on $v_1v_2$. Same argument applies when (a)--(e) in Condition B are satisfied. Therefore, the identifiability of $l_i$ (with end-points $v_1$ and $v_2$) of Category 2.1 can be determined by Algorithm~\ref{Alg:triangleC2}.
\hfill$\blacksquare$

\subsection{Proof of Lemma~\ref{lemma:Alg4}}
\begin{figure}[tb]
\centering
\includegraphics[width=2.4in]{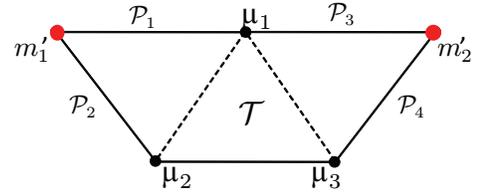}
\vspace{-.5em}
\caption{Link identifications of Category 2.3.}\label{fig:Category4}
\vspace{-.5em}
\end{figure}
Let the two conjugate pairs be $\{\mu_1,\mu_2\}$ and $\{\mu_1,\mu_3\}$ and $S_1$ ($S_2$) the set of immediately neighboring triconnected components connecting to $\mathcal{T}$ via $\{\mu_1,\mu_2\}$ ($\{\mu_1,\mu_3\}$), as illustrated in Fig.~\ref{fig:Category4}. Since $\{\mu_1,\mu_2\}$ and $\{\mu_1,\mu_3\}$ are 2-vertex-cuts, we know that there exist internally vertex disjoint paths $\mathcal{P}_1$, $\mathcal{P}_2$, $\mathcal{P}_3$ and $\mathcal{P}_4$. If $\mu_1\mu_2$ (or $\mu_1\mu_3$) is a real link, then $\mu_1\mu_2$ (or $\mu_1\mu_3$) is known as a Cross-link \cite{Ma13IMC} since $\mu_2\mu_3$ can be replaced by a path in neighboring biconnected components if $\mu_2\mu_3$ is virtual. Therefore, $\mu_1\mu_2$ (or $\mu_1\mu_3$) is identifiable. Now we focus on the identification of $\mu_2\mu_3$ in $\mathcal{T}$ (when $\mu_2\mu_3$ is a real link). The conditions to guarantee the identifiability of $\mu_2\mu_3$ is:
\\(link $\mu_1\mu_2$ is real \textbf{OR} $|S_1|\geq 2$ \textbf{OR} one component in $S_1$ is 3-vertex-connected) \textbf{AND} (link $\mu_1\mu_3$ is real \textbf{OR} $|S_2|\geq 2$ \textbf{OR} one component in $S_2$ is 3-vertex-connected).
\\Suppose the above condition is not satisfied. Then we can prove $\mu_2\mu_3$ is unidentifiable as follows.

\begin{figure}[tb]
\centering
\includegraphics[width=3.5in]{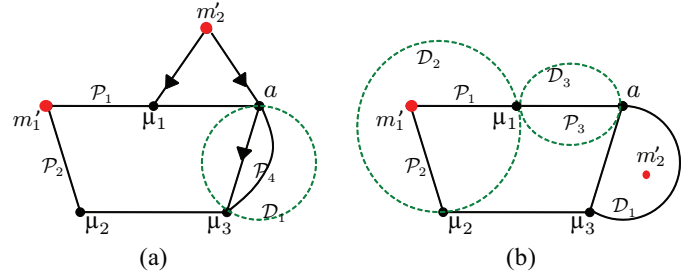}
\vspace{-.5em}
\caption{Category 2.3 - unidentifiable $\mu_2\mu_3$.}\label{fig:Category4Unidentifiable}
\vspace{-.5em}
\end{figure}

We first consider the condition: link $\mu_1\mu_3$ is real \textbf{OR} $|S_2|\geq 2$ \textbf{OR} one component in $S_2$ is 3-vertex-connected. If not satisfied, then it implies that there is no real link $\mu_1\mu_3$ and the only immediately neighboring triconnected component is a triangle, i.e., $\mu_1$-$a$-$\mu_3$ in Fig.~\ref{fig:Category4Unidentifiable} ($\mu_1a$ and $\mu_3a$ can be virtual links as well).

1) If agent $m'_2$ is in the location shown in Fig.~\ref{fig:Category4Unidentifiable}(a), then all paths from $m'_1$ to $m'_2$ traversing $\mu_2\mu_3$ must use one simple path in $\mathcal{D}_1$. Therefore, the best case is that we can compute the sum metric of link ${\mu_2\mu_3}$ and another link which is incident to $\mu_3$ in $\mathcal{D}_1$, but cannot compute them separately.

2) If agent $m'_2$ is in the location shown in Fig.~\ref{fig:Category4Unidentifiable}(b), then $\mathcal{P}_3$ and $\mu_2\mu_3$ become a ``double bridge'' connecting $\mathcal{D}_2$ and $\mathcal{D}_1$. Abstracting $\mathcal{P}_3$ as a single link, \cite{Ma13IMC} proves that none of the links in a double bridge is identifiable when constraining the measurement paths to simple paths. If we choose other paths as $\mathcal{P}_3$ in $\mathcal{D}_3$, then the same argument applies. Therefore, based on 1) and 2), $\mu_2\mu_3$ is unidentifiable.

Analogously, we can prove that $\mu_2\mu_3$ is unidentifiable when condition (link $\mu_1\mu_2$ is real \textbf{OR} $|S_1|\geq 2$ \textbf{OR} one component in $S_1$ is 3-vertex-connected) is not satisfied.

When the required conditions are satisfied, we can prove that $\mu_2\mu_3$ is identifiable as follows:

\vspace{-.5em}
If $\mu_1\mu_2$ ($\mu_1\mu_3$) is a virtual link, then it can be replaced by a path in a neighboring component. For instance, if $|S_1|\geq 2$, then one replacement path can be found in one component of $S_1$. If one component in $S_1$ is 3-vertex-connected, then there exist 2 internally vertex disjoint paths (each with the order greater than 1) connecting $\mu_1$ and $\mu_2$. Thus, we can choose one of them as a replacement path. Note that the virtual links possibly involved in the replacement paths can be further replaced by the paths in their neighboring components within the same parent biconnected component. After these replacement operations, $\mu_2\mu_3$ in Fig.~\ref{fig:Category4} is a Shortcut (defined in \cite{Ma13IMC}), which is proved to be identifiable in \cite{Ma13IMC}.

Therefore, Algorithm~\ref{Alg:triangleC4} can determine all identifiable/unidentifiable links in a triangle triconnected component of Category 2.3.
\hfill$\blacksquare$

\subsection{Proof of Lemma~\ref{lemma:Alg5}}
For direct link $\mu_1\mu_2$ in $\mathcal{T}$, if there exist neighboring triconnected components connecting to $\mathcal{T}$ by 2-vertex cut $\{\mu_1,\mu_2\}$ \textbf{AND} $\mu_1$ and $\mu_2$ are not the (only two) vantages in one of these neighboring components $\mathcal{T}_N$, then $\mu_1\mu_2$ is not a direct link connecting two vantages in $\mathcal{T}_N$. Therefore, if $\mathcal{T}_N$ is of Category 1 or 2, then Algorithm~DAIL can determine the identifiability of $\mu_1\mu_2$; if $\mathcal{T}_N$ is of Category 3, then $\mu_1\mu_2$ is identifiable since it is not a direct link in $\mathcal{T}_N$. The above argument is complete, since the number of vantages in $\mathcal{T}_N$ cannot be less than 2 as the parent biconnected component of $\mathcal{T}$ (and $\mathcal{T}_N$) contains at least 2 agents\footnote{In Algorithm~2, we only consider the biconnected components with 2 or more agents.}.

\begin{figure}[tb]
\centering
\includegraphics[width=2.7in]{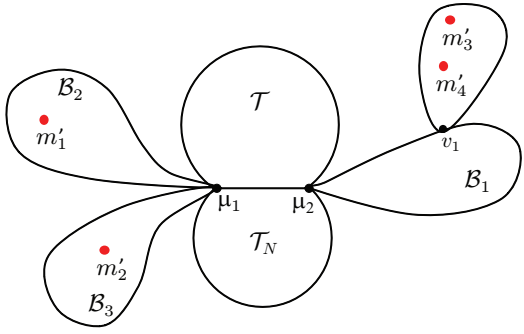}
\vspace{-.5em}
\caption{Direct link $\mu_1\mu_2$ identification.}\label{Fig:directLinkIdenti}
\vspace{-.5em}
\end{figure}

\begin{figure}[tb]
\centering
\includegraphics[width=1.7in]{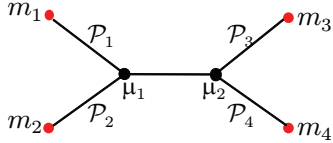}
\vspace{-.9em}
\caption{Direct link $\mu_1\mu_2$ identification in the abstracted graph.} \label{Fig:DirectLinkIdentification}
\vspace{-.5em}
\end{figure}

If the condition in line~1 of Algorithm~B is not satisfied, then there is no neighboring triconnected component connecting to $\mathcal{T}$ by 2-vertex cut $\{\mu_1,\mu_2\}$, or all these neighboring components contain only $\mu_1$ and $\mu_2$ as vantages, shown in Fig.~\ref{Fig:directLinkIdenti}. In this case, the identifiability of $\mu_1\mu_2$ can be determined with the assistance of neighboring biconnected components. Let $\mathcal{B}_1$ be the only biconnected component connecting to $\mathcal{T}$ via $\mu_2$, as shown in Fig.~\ref{Fig:directLinkIdenti}. Suppose $\mu_2$ is not a real monitor and $v_1$ is the only agent (except for $\mu_2$) in $\mathcal{B}_1$. Then we can at most identify the sum metric of $\mu_1\mu_2$ and a path connecting $\mu_2$ and $v_1$ in $\mathcal{B}_1$, but not the individual link metric on $\mu_1\mu_2$. Therefore, if $\mathcal{B}_1$ is the only biconnected component connecting to $\mathcal{T}$ by $\mu_2$ and $\mu_2$ is not a real monitor, then $\mathcal{B}_1$ must contain at least 2 agents (except for $\mu_2$) for identifying $\mu_1\mu_2$. The same argument applies to the biconnected components connecting to $\mathcal{T}$ via $\mu_1$. Suppose none of $\mu_1$ or $\mu_2$ is a real monitor. If there are more than one biconnected component connecting to $\mathcal{T}$ via $\mu_1$ (or $\mu_2$), e.g., $\mathcal{B}_2$ and $\mathcal{B}_3$ in Fig.~\ref{Fig:directLinkIdenti}, then the total number of agents (except for $\mu_1$ (or $\mu_2$)) must be no less than 2 for identifying $\mu_1\mu_2$. If these required conditions regarding neighboring biconnected components of $\mu_1$ and $\mu_2$ are satisfied, then this scenario can be abstracted as Fig.~\ref{Fig:DirectLinkIdentification}, which satisfies the condition (suppose $\mu_1$ and $\mu_2$ are not monitors) in Theorem III.3, and thus $\mu_1\mu_2$ is identifiable. Note that for node $\mu_1$, if $\mu_1$ is a real monitor, then there is no requirement for the neighboring biconnected components of $\mu_1$, since the subgraph containing $\mu_1\mu_2$, $\mathcal{P}_3$, $\mathcal{P}_4$ and their associated end-points also satisfies the condition in Theorem III.3. The same argument applies to the case when $\mu_2$ is a real monitor. Consequently, Algorithm~B can determine the identifiability of $\mu_1\mu_2$ when the condition in line~1 is not satisfied, whereas the case satisfying the condition in line~1 can be determined in neighboring triconnected components (see line~2 of Algorithm~B).
\hfill$\blacksquare$

\subsection{Proof of Theorem~\ref{theorem:uniqueTriNum}}
\begin{definition}
\label{def:merge}
\emph{merge}: In graph $\mathcal{G}$, two triconnected components $\mathcal{T}_1$ and $\mathcal{T}_2$ with a common virtual link $ab$, i.e., $ab\in L(\mathcal{T}_1)\cap L(\mathcal{T}_2)$, can be merged. The resulting graph after the merging operation is $\mathcal{G}=(V',L')$, where $V'=V(\mathcal{T}_1)\cup V(\mathcal{T}_2)$, $L'=L(\mathcal{T}_1)\cup L(\mathcal{T}_2)\setminus \{ab\}$.
\end{definition}

\emph{1)} In \cite{Hopcroft73}, it has proved (see Lemma~2 in \cite{Hopcroft73}) that 3-vertex-connected and single-link components in a graph decomposition are unique. Moreover, Lemma~2 in \cite{Hopcroft73} also proved that if we merge (see Definition~\ref{def:merge}) all triangle components as much as possible, then a set of polygons are obtained and these polygons are unique. Since there are multiple ways to decompose a polygon, the generated triangle components may not be unique.

\begin{figure}[tb]
\centering
\includegraphics[width=3.5in]{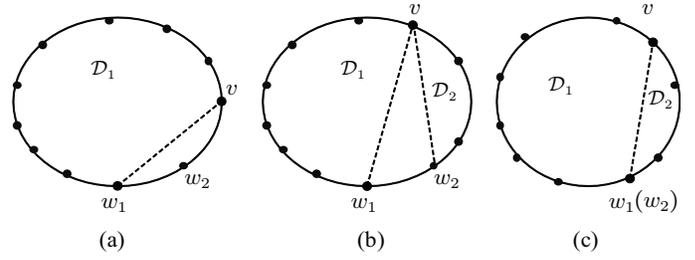}
\vspace{-.9em}
\caption{Decomposition of triconnected components in a polygon.} \label{Fig:triconnectedNumber}
\vspace{-.5em}
\end{figure}

\emph{2)} To prove the total number of triconnected components $\mathcal{N}_{\mathcal{T}}$ in $\mathcal{G}$ is fixed, it suffices to show that the total number of triangle components is fixed regardless of the mechanisms being used for graph decomposition, since all other types of triconnected components are unique as proved in \emph{1)}. To this end, we have one claim: for a unique polygon $\mathcal{D}$ obtained in \emph{1)} with $\tau$ nodes ($\tau\geq 3$), the number of triangles after graph decomposition on $\mathcal{D}$ is always $\tau-2$, which is irrespective of decomposition strategies. We prove this claim by induction. (1) If $\tau=3$, then there is no need to further decompose $\mathcal{D}$, since $\mathcal{D}$ is already a triangle. Hence, in this case, the number of triangles is only one, i.e., $\tau-2$. (2) Suppose the claim is true for $\tau=k$ ($k$ is an integer, $k\geq 3$). (3) Consider the case that $\tau=k+1$. Fig.~\ref{Fig:triconnectedNumber}(a)--(b) display a polygon with $k+1$ nodes. For real link $w_1w_2$ on this polygon, after decomposition, it must belong to one triangle $\mathcal{T}$. Let the third node on $\mathcal{T}$ be $v$ (see Fig.~\ref{Fig:triconnectedNumber}(a)--(b)). Then there are two cases for $\mathcal{T}$: (i) $vw_2$ (or $vw_1$) is a real link in the original graph (Fig.~\ref{Fig:triconnectedNumber}(a)). In this case, excluding triangle $\mathcal{T}$, the remaining graph is polygon $\mathcal{D}_1$ with virtual link $vw_1$ (or $vw_2$). Then to get all triangles, $\mathcal{D}_1$ needs to be further decomposed. We know $|V(\mathcal{D}_1)|=k$, and thus it corresponds to $k-2$ triangles according to the hypothesis in (2). Therefore, the total number of triangles is $(k+1)-2$; (ii) neither $vw_1$ nor $vw_2$ is a real link in the original graph $\mathcal{D}$. Then excluding triangle $\mathcal{T}$, the remaining graphs are $\mathcal{D}_1$ and $\mathcal{D}_2$ (Fig.~\ref{Fig:triconnectedNumber}(b)). To further decompose $\mathcal{D}_1$ and $\mathcal{D}_2$, we can first combine $\mathcal{D}_1$ and $\mathcal{D}_2$ as follows: Let $w_1=w_2$. The graph combination $\mathcal{D}'$ is the union of $\mathcal{D}_1$ and $\mathcal{D}_2$, excluding virtual link $vw_1$ ($vw_2$) (see Fig.~\ref{Fig:triconnectedNumber}(c)). Then the decomposition of $\mathcal{D}_1$ and $\mathcal{D}_2$ is equivalent to the decomposition of $\mathcal{D}'$, conditioned on that the vertex-cut $\{v,w_1\}$ (the same as $\{v,w_2\}$) must be used. We know the generated $\mathcal{D}'$ is a polygon, which contains $k$ nodes, and thus it always corresponds to $k-2$ triangles (including the case that vertex-cut $\{v,w_1\}$ ($\{v,w_2\}$) must be used). Therefore, original graph $\mathcal{D}$ corresponds to $(k+1)-2$ triangles, completing the proof.
\hfill$\blacksquare$

\subsection{Proof of Theorem~\ref{theorem:FLIcorrectness}}

In Theorem III.2 and III.3, the prerequisite for network identifiability is that all involved links can be used for constructing measurement paths. In DAIL, we sequentially consider each triconnected component which possibly contains virtual links. These virtual links, however, do not exist in real networks. To tackle with this issue, Claim 1 states that a replacement path can be found for a virtual link whose end-points $\{v_1,v_2\}$ form a Type-0-VC wrt $\mathcal{T}$. We will discuss the virtual links, not covered by Claim 1, in the following link identifications for the three (where there are four sub-categories for Category 2) triconnected component categories.

\begin{figure}[tb]
\centering
\includegraphics[width=3.1in]{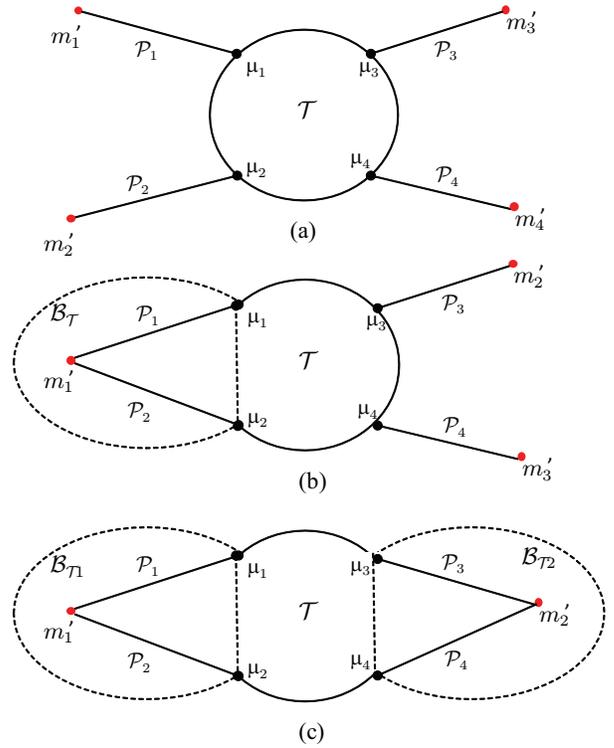}
\vspace{-.5em}
\caption{Link identifications of Category 1.}\label{fig:Category1}
\vspace{-.5em}
\end{figure}

\emph{1)} Category 1.

It suffices to only consider the case of a triconnected component $\mathcal{T}$ with 4 vantages. This is because, for a triconnected component $\mathcal{T}$ with more than 4 vantages, only four of them are useful for identifying $\mathcal{T}$. Since $|V(\mathcal{T})|\geq 4$, $\mathcal{T}$ must be 3-vertex-connected, i.e., not a triangle nor a bond. There are 3 possible scenarios after replacing all possible virtual links (according to Claim 1) in $\mathcal{T}$ by the corresponding real paths in neighboring components. (i) For the four vantages, each vantage-to-agent path is within a different biconnected component: As illustrated in Fig.~\ref{fig:Category1}(a), there exist 4 vertex disjoint vantage-to-agent paths $\mathcal{P}_1,\ldots,\mathcal{P}_4$. Therefore, abstracting $\mathcal{P}_1,\ldots,\mathcal{P}_4$ as single links, the graph in Fig.~\ref{fig:Category1}(a) satisfies the condition in Theorem~III.3, and thus $\mathcal{T}$ is fully identifiable. (ii) For two vantages in $\{\mu_1,\ldots,\mu_4\}$, say $\mu_1$ and $\mu_2$, their vantage-to-agent paths are within the same biconnected component, as shown in Fig.~\ref{fig:Category1}(b). In this case, there exists at least one agent ($m'_1$) in the neighboring biconnected component $\mathcal{B}_{\mathcal{T}}$, and two internally vertex disjoint paths $\mathcal{P}_1$ and $\mathcal{P}_2$ in Fig.~\ref{fig:Category1}(b). Hence, abstracting $\mathcal{P}_1,\ldots,\mathcal{P}_4$ in Fig.~\ref{fig:Category1}(b) as single links, the graph in Fig.~\ref{fig:Category1}(b) also satisfies the condition in Theorem~III.3 even if $\mu_1\mu_2$ is a virtual link, and thus $\mathcal{T}$ is fully identifiable. (iii) For two pairs of vantages in $\{\mu_1,\ldots,\mu_4\}$, the vantage-to-agent paths for each pair are within the same biconnected component, as shown in Fig.~\ref{fig:Category1}(c). Following the similar argument in (ii), the graph in Fig.~\ref{fig:Category1}(c) satisfies the condition in Theorem~III.3 even if $\mu_1\mu_2$ and $\mu_3\mu_4$ are virtual links as $\mathcal{T}$ is 3-vertex-connected, and thus $\mathcal{T}$ is fully identifiable. On top of Fig.~\ref{fig:Category1}(c), there may exist other pairs of vantages, whose vantage-to-agent paths are within the same biconnected component. In this case, these vantage-to-agent paths can be ignored, since $\mathcal{P}_1,\ldots,\mathcal{P}_4$ in Fig.~\ref{fig:Category1}(c) are already sufficient to identify $\mathcal{T}$.

\emph{2)} Category 2.1.

\begin{figure}[tb]
\centering
\includegraphics[width=3.2in]{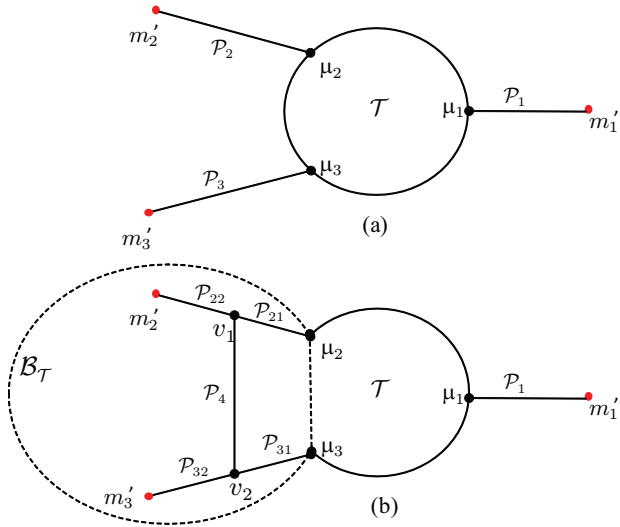}
\vspace{-.5em}
\caption{Link identifications of Category 2.1.}\label{fig:Category2}
\vspace{-.5em}
\end{figure}

Suppose $\mathcal{T}$ of Category 2.1 (Fig.~\ref{fig:Category2}) is 3-vertex-connected, and all possible virtual links in $\mathcal{T}$ are replaced by the corresponding real paths in neighboring components of $\mathcal{T}$. There are two possible scenarios. (i) For the three vantages, each vantage-to-agent path is within a different biconnected component: As illustrated in Fig.~\ref{fig:Category2}(a), there exist 3 vertex disjoint vantage-to-agent paths $\mathcal{P}_1,\ldots,\mathcal{P}_3$. Therefore, abstracting $\mathcal{P}_1,\ldots,\mathcal{P}_3$ as single links, the graph in Fig.~\ref{fig:Category2}(a) satisfies the condition in Theorem~III.3, and thus $\mathcal{T}$ is fully identifiable. (ii) For two vantages in $\{\mu_1,\ldots,\mu_3\}$, say $\mu_2$ and $\mu_3$, their vantage-to-agent paths are within the same biconnected component $\mathcal{B}_{\mathcal{T}}$, as shown in Fig.~\ref{fig:Category2}(b). In this case, there exist path $\mathcal{P}_4$ connecting $v_1$ and $v_2$, and $v_1$ ($v_2$) has at least 3 neighbors in $\mathcal{B}_{\mathcal{T}}$. Moreover, $\mu_2$ and $\mu_3$ must have at least two neighbors in $\mathcal{T}$ (we have assumed all other virtual links in $\mathcal{T}$ have been replaced by neighboring components), since $\mathcal{T}$ is 3-vertex-connected. Therefore, $\mu_2$ and $\mu_3$ are not nodes with only two neighbors, and thus they do not have to be monitors for complete network identification of Fig.~\ref{fig:Category2}(b) (see the explanation of MMP in \cite{MaInfocom14sub}). Abstracting $\mathcal{P}_1,\mathcal{P}_{21},\mathcal{P}_{22},\mathcal{P}_{31},\mathcal{P}_{32},\mathcal{P}_4$ in Fig.~\ref{fig:Category2}(b) as single links, the graph in Fig.~\ref{fig:Category2}(b) satisfies the condition in Theorem~III.3 even if $\mu_2\mu_3$ is a virtual link, and thus 3-vertex-connected component $\mathcal{T}$ is fully identifiable. Meanwhile, same argument applies if $\{\mu_1,\mu_2\}$ and/or $\{\mu_1,\mu_3\}$ are type-$k$-VCs ($k\geq 2$).

If $\mathcal{T}$ is a triangle, then the identifiability of $\mathcal{T}$ is determined by Lemma~\ref{lemma:Alg3}.

\begin{figure}[tb]
\centering
\includegraphics[width=3in]{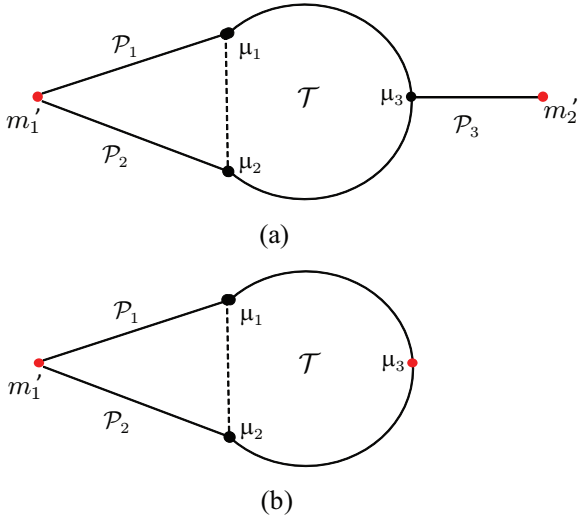}
\vspace{-.5em}
\caption{Link identifications of Category 2.2.}\label{fig:Category3}
\vspace{-.5em}
\end{figure}

\emph{3)} Category 2.2.

Fig.~\ref{fig:Category3} illustrates the case of Category 2.2, where $\mu_1$, $\mu_2$ and $\mu_3$ are the vantages in $\mathcal{T}$ and $\{\mu_1,\mu_2\}$ is a conjugate pair. Since the parent biconnected component of $\mathcal{T}$ is 2-vertex-connected, there exist two paths $\mathcal{P}_1$ and $\mathcal{P}_2$ connecting to an agent $m'_1$. $\mathcal{P}_1$ and $\mathcal{P}_2$ are internally vertex disjoint, since if $\mathcal{P}_1$ and $\mathcal{P}_2$ must have a common node (except $m'_1$), then this common node is a cut-vertex, contradicting the property of biconnectivity (see Lemma~\ref{lemma:vDisjointPaths}). Moreover, there exists path $\mathcal{P}_3$ connecting $\mu_3$ and $m'_2$ with $\mathcal{P}_1\cap\mathcal{P}_3=\emptyset$ and $\mathcal{P}_2\cap\mathcal{P}_3=\emptyset$. By Claim 1, all virtual links in $\mathcal{T}$ (except $\mu_1\mu_2$) can be replaced by the corresponding real paths in neighboring components; therefore, we only need to consider one virtual link $\mu_1\mu_2$ (if any) in Fig.~\ref{fig:Category3}. In the proof of Lemma.~\ref{lemma:effectiveMonitor}, Fig.~\ref{fig:EffectiveMonitorProof} illustrates that, using two monitors, if one monitor has only one neighbor $v$, then all neighboring links of $v$ are unidentifiable. Therefore, in the case of employing two monitors, this neighboring node of each monitor $v$ is an effective monitor. We can therefore abstract $m'_1$ and $\mu_3$ as two monitors (by Lemma~\ref{lemma:effectiveMonitor}) and $\mathcal{P}_1$ and $\mathcal{P}_2$ as single links in Fig.~\ref{fig:Category3}(a) to obtain Fig.~\ref{fig:Category3}(b). Fig.~\ref{fig:Category3}(b) satisfies the interior graph identifiability conditions (Theorem III.2) and exterior link unidentifiability conditions (Theorem III.1) naturally if there exists real link $\mu_1\mu_2$. Now we consider that there is no link $\mu_1\mu_2$ in the original graph, i.e., $\mu_1\mu_2$ is a virtual link. If $\mathcal{T}$ is a triangle, then the two exterior links ($\mu_1\mu_3$ and $\mu_2\mu_3$ if any) in Fig.~\ref{fig:Category3}(b) are unidentifiable according to Theorem III.1. Now consider the case that $\mathcal{T}$ is 3-vertex-connected. Deleting any two links\footnote{The link can be a path from neighboring biconnected component for virtual link replacement in $\mathcal{T}$.} in $\mathcal{T}$, the resulting graph is connected as 3-vertex-connectivity of $\mathcal{T}$ implies 3-edge-connectivity. Deleting $\mathcal{P}_1$ (or $\mathcal{P}_2$) and one link in $\mathcal{T}$, we also get a connected remaining graph. Thus, Fig.~\ref{fig:Category3}(b) satisfies Condition \textcircled{\small 1} \normalsize in Theorem III.2. Now consider deleting some vertices in $\mathcal{T}$. Deleting any two vertices in $\mathcal{T}$, the remaining graph of $\mathcal{T}$ is still connected as $\mathcal{T}$ is 3-vertex-connected. In this case, if $m'_1$ is isolated ($\mu_1$ and $\mu_2$ are deleted), $m'_1$ can reconnect to the remaining part of $\mathcal{T}$ by added link $m'_1\mu_3$ (see Condition \textcircled{\small 2} \normalsize in Theorem III.2). When deleting $m'_1$ and a node in $\mathcal{T}$, the remaining graph of Fig.~\ref{fig:Category3}(b) is obviously connected, thus satisfying Condition \textcircled{\small 2} \normalsize in Theorem III.2. Therefore, for $\mathcal{T}$ of Category 2.2, all real links incident to $\mu_3$ are unidentifiable and the remaining links are identifiable.

\emph{4)} Category 2.3.

Category 2.3 is illustrated in Fig.~\ref{fig:Category4}. Due to the 2-vertex-connectivity of the parent biconnected component of $\mathcal{T}$, there exist pairwise internally vertex disjoint vantage-to-agent paths $\mathcal{P}_1$, $\mathcal{P}_2$, $\mathcal{P}_3$ and $\mathcal{P}_4$. Consider the case that $\mathcal{T}$ is 3-vertex-connected and all virtual links in $\mathcal{T}$ (except $\mu_1\mu_2$ and $\mu_1\mu_3$ if they exist) are replaced by the corresponding real paths in neighboring components by Claim 1. Then abstracting $\mathcal{P}_1,\ldots,\mathcal{P}_4$ as single links, the graph in Fig.~\ref{fig:Category4} satisfies the condition in Theorem III.2 even if $\mu_1\mu_2$ or $\mu_1\mu_3$ does not exist (i.e., virtual links). Therefore, all interior links in Fig.~\ref{fig:Category4} (i.e., all links in $\mathcal{T}$) are identifiable. In the case that $\mathcal{T}$ is a triangle, the identifiability of $\mathcal{T}$ is determined by Lemma~\ref{lemma:Alg4}.

\emph{5)} Category 2.4.

\begin{figure}[tb]
\centering
\includegraphics[width=1.5in]{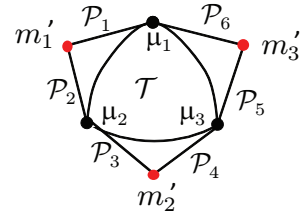}
\vspace{-.5em}
\caption{Link identifications of Category 2.4.}\label{fig:Category5}
\vspace{-.5em}
\end{figure}

Fig.~\ref{fig:Category5} illustrates triconnected component of Category~2.4. There exist pairwise internally vertex disjoint vantage-to-agent paths $\mathcal{P}_1,\ldots,\mathcal{P}_6$. Replacing all virtual links in $\mathcal{T}$ (except $\mu_1\mu_2$, $\mu_2\mu_3$ and $\mu_1\mu_3$ if they exist) by the corresponding real paths in neighboring components by Claim 1 and abstracting $\mathcal{P}_1,\ldots,\mathcal{P}_6$ as single links, the resulting graph in Fig.~\ref{fig:Category5} satisfies the condition in Theorem~III.3 even if $\mu_1\mu_2$, $\mu_2\mu_3$ or $\mu_1\mu_3$ does not exist (i.e., virtual links). Therefore, all links in $\mathcal{T}$ of Category~2.4 are identifiable.

\emph{6)} Category 3.

\begin{figure}[tb]
\centering
\includegraphics[width=2.3in]{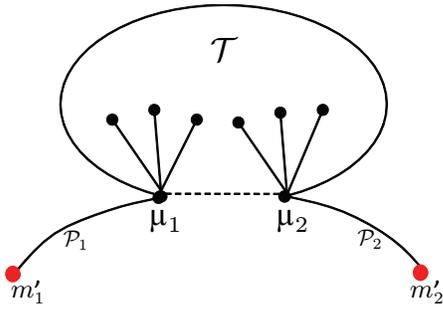}
\vspace{-.5em}
\caption{Link identification of Category 3.}\label{fig:Category6}
\vspace{-.5em}
\end{figure}

Suppose $\mathcal{T}$ contains vantages $\mu_1$ and $\mu_2$, as illustrated Fig.~\ref{fig:Category6}. There exist vertex disjoint paths $\mathcal{P}_1$ and $\mathcal{P}_2$ in the parent biconnected components. Thus, it is equivalent to the case of Fig.~\ref{fig:EffectiveMonitorProof}. By Claim 1, all virtual links in $\mathcal{T}$ (except $\mu_1\mu_2$) can be replaced by the corresponding real paths in neighboring components; therefore, for $\mathcal{T}$, all real links incident to $\mu_1$ and $\mu_2$ are unidentifiable and the remaining links are identifiable. Moreover, the identifiability of $\mu_1\mu_2$ (if it exists) is determined by Lemma~\ref{lemma:Alg5}.

In DAIL, when triconnected component $\mathcal{T}$ is of Category 2, satisfying the conditions in lines~5 and 6 (of DAIL), then this is actually the case of Category 2.2, which is processed by line~7 (in DAIL). For the case that 3-vertex-connected $\mathcal{T}$ is of Category 2.1, 2.3 and 2.4, the above proof shows that all links in $\mathcal{T}$ are identifiable (processed by line 9 in DAIL). When a Category 2 component is a triangle, the identifiability of this component is determined by Algorithm~A (in line 12 of DAIL).

Given the fact that the triconnected component decomposition is not unique, finally, we prove that link identifiability determined by DAIL does not depend on the mechanism for triconnected component decomposition. According to Theorem~\ref{theorem:uniqueTriNum}, all triconnected components are unique except for the triangle components. Therefore, for a link $l$ in $\mathcal{G}$, under a given graph decomposition, if $l$ falls into a 3-vertex-connected or single-link component $\mathcal{T}$, then $l$ is always in this component $\mathcal{T}$ under all other decomposition mechanisms, since $\mathcal{T}$ is unique. Hence, it suffices to show that the identifiability of links within triangle components remains the same for all possible graph decompositions. Consider link $l$ within a triangle component $\mathcal{T}$ under one graph decomposition. If $\mathcal{T}$ contains 3 real links, then $\mathcal{T}$ is also unique under all possible graph decompositions (see Lemma 2 in \cite{Hopcroft73}). Now Suppose $\mathcal{T}$ contains virtual links. Then there are 5 cases.

\begin{figure}[tb]
\centering
\includegraphics[width=2.6in]{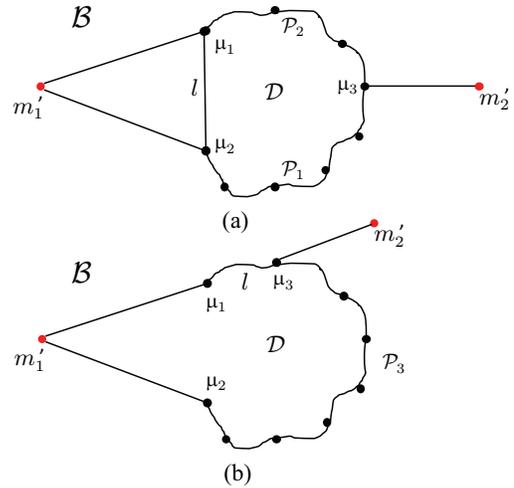}
\vspace{-.5em}
\caption{Link identification of Category 2.2 in different triconnected components.}\label{fig:triangle22}
\vspace{-.5em}
\end{figure}

\begin{figure}[tb]
\centering
\includegraphics[width=3.4in]{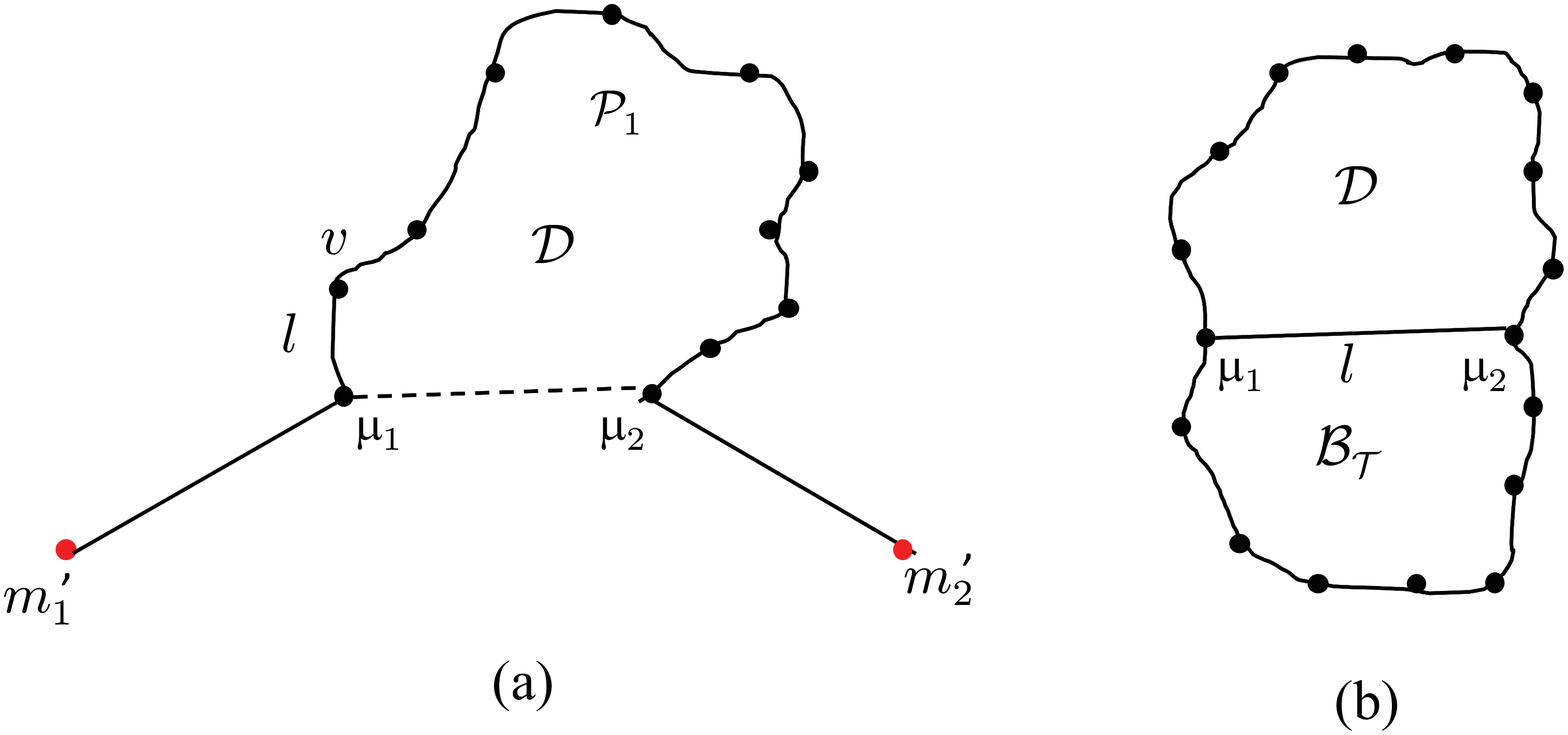}
\vspace{-.5em}
\caption{Link identification of Category 3 in different triconnected components.}\label{fig:triangleCategory3}
\vspace{-.5em}
\end{figure}

\begin{enumerate}[(i)]
  \item $\mathcal{T}$ is of Category 2.1. In this case, $\mathcal{T}$ is identified by Algorithm~\ref{Alg:triangleC2} (the correctness is shown in Lemma~\ref{lemma:Alg3}). Throughout the proof of Lemma~\ref{lemma:Alg3}, we consider the identification of $l$ ($l\in L(\mathcal{T})$) only in its parent biconnected component, instead of triconnected component $\mathcal{T}$, e.g., $v_1v_2$ in Fig.~\ref{Fig:quasiBC}(a) can be classified into multiple triangle components; however, its parent biconnected component is unique. Therefore, the generated triconnected component $\mathcal{T}$ involving $l$ under different decomposition mechanisms does not affect the identifiability of $l$;
  \item $\mathcal{T}$ is of Category 2.2. This case is shown in Fig.~\ref{fig:triangle22}. Note all links in Fig.~\ref{fig:triangle22} are real links. If $l=\mu_1\mu_2$ as shown in Fig.~\ref{fig:triangle22}(a), then within the parent biconnected component, there exist real paths $\mathcal{P}_1$ and $\mathcal{P}_2$, connecting $\mu_3$ to $\mu_2$ and $\mu_1$, respectively. Abstracting $\mathcal{P}_1$ and $\mathcal{P}_2$ as single links, $l$ is always identifiable since $l$ is a cross-link \cite{Ma13IMC}. When $\mathcal{D}$ in Fig.~\ref{fig:triangle22}(a) is a polygon, there are multiple triangle decompositions for $\mathcal{D}$; nevertheless, the detailed triangle decomposition does not prevent $l$ in Fig.~\ref{fig:triangle22}(a) from being a cross-link. If $l=\mu_1\mu_3$ (or $l=\mu_2\mu_3$) as shown in Fig.~\ref{fig:triangle22}(b), then similar to Fig.~\ref{fig:triangle22}(a), within the parent biconnected component, there exists path $\mathcal{P}_3$, connecting $\mu_2$ to $\mu_3$ (Fig.~\ref{fig:triangle22}(b)). Abstracting $\mathcal{P}_3$ as a single link, both $l$ and $\mathcal{P}_3$ are exterior links, thus unidentifiable. It is possible that $\mathcal{D}$ in Fig.~\ref{fig:triangle22}(b) is a polygon, then there are multiple triangle decompositions for $\mathcal{D}$; nevertheless, the detailed triangle decomposition does not affect the identifiability of $l$ in Fig.~\ref{fig:triangle22}(b);
  \item $\mathcal{T}$ is of Category 2.3. In this case, $\mathcal{T}$ is identified by Algorithm~\ref{Alg:triangleC4} (the correctness is shown in Lemma~\ref{lemma:Alg4}). Throughout the proof of Lemma~\ref{lemma:Alg4}, we consider the identification of $l$ ($l\in L(\mathcal{T})$) only in its parent biconnected component, instead of triconnected component $\mathcal{T}$, e.g., $\mu_2\mu_3$ in Fig.~\ref{fig:Category4Unidentifiable}(b) can be classified into multiple triangle components; however, its parent biconnected component is unique. Therefore, the generated triconnected component $\mathcal{T}$ involving $l$ under different decomposition mechanisms does not affect the identifiability of $l$;
  \item $\mathcal{T}$ is of Category 2.4. This case is shown in Fig.~\ref{fig:Category5}, where each identifiable link $\mu_1\mu_2$, $\mu_1\mu_3$, or $\mu_2\mu_3$ is a cross-link. The property of being a cross-link does not depend on how triconnected component $\mathcal{T}$ is generated; therefore, the identifiability of $l$ within $\mathcal{T}$ remains the same for all possible graph decompositions;
  \item $\mathcal{T}$ is of Category 3. This case is shown in Fig.~\ref{fig:triangleCategory3}, where $\mathcal{D}$ is 2-vertex-connected and all links are real except that $\mu_1\mu_2$ possibly is a virtual link. In Fig.~\ref{fig:triangleCategory3}(a), if $l=v\mu_1$, then $l$ is always unidentifiable irrespective of the graph decomposition of $\mathcal{D}$ ($\mathcal{D}$ is a polygon). If $l=\mu_1\mu_2$ as shown in Fig.~\ref{fig:triangleCategory3}(b), then the identifiability of $l$ is determined by neighboring components according to Algorithm B. Within the same parent biconnected component, let $\mathcal{B}_\mathcal{T}$ be the neighboring biconnected component connecting to $\mathcal{D}$ by $\{\mu_1,\mu_2\}$ (see Fig.~\ref{fig:triangleCategory3}(b)). In $\mathcal{B}_\mathcal{T}$, for a triconnected component involving $l$, if this triconnected component is of Category 1, 2.1, 2.2, 2.3, or 2.4, then the arguments in (i)--(iv) and Theorem~\ref{theorem:uniqueTriNum} have proved the identifiability of $l$ is independent of the graph decomposition of $\mathcal{B}_\mathcal{T}$; however, if this triconnected component is of Category 3, then the identifiability of $l$ in Fig.~\ref{fig:triangleCategory3}(b) relies on the external agent connections to neighboring biconnected components according to Algorithm B, and thus the identifiability of $l$ only depends on the biconnected component decomposition, which is unique;
\end{enumerate}

Consequently, with the complete coverage of three categories, the identification efficacy of each category, and independence to graph decompositions, DAIL can determine all identifiable/unidentifiable links.
\hfill$\blacksquare$

\subsection{Proof of Theorem~\ref{theorem:candidates}}
Let $d^\mathcal{T}(v):=|\mathcal{A}^\mathcal{T}(v)|$, $\mathcal{U}_{v_1 v_2}:=\mathcal{A}^\mathcal{T}(v_1)\cup \mathcal{A}^\mathcal{T}(v_2)$, and $d^\mathcal{T}_{m}:=\max d^\mathcal{T}(v)$, $v\in \{v_1,v_2,v_3\}$.

\emph{1)} Suppose $|\mathcal{T}|\geq 3$. We first prove that $d^\mathcal{T}(w)\geq d^\mathcal{T}_{m}$ for $w\in V(\mathcal{T})\setminus \{v_1,v_2,v_3\}$. According to lines~3 and 5 in Algorithm~3 of \cite{MaInfocom14sub}, we have $d^\mathcal{T}(v_1)+ d^\mathcal{T}(v_2)-1\leq d^\mathcal{T}(v_1)+ d^\mathcal{T}(v_3)$, and thus $d^\mathcal{T}(v_2)\leq d^\mathcal{T}(v_3)+1$. There are two possible cases for the relationship between $d^\mathcal{T}(v_2)$ and $d^\mathcal{T}(v_3)$. (i) $d^\mathcal{T}(v_2)\geq d^\mathcal{T}(v_3)$: Then $d^\mathcal{T}(v_3)\leq d^\mathcal{T}(v_2)\leq d^\mathcal{T}(v_3)+1$. Accordingly, we have $d^\mathcal{T}(v_2)- d^\mathcal{T}(v_3)=1$ or $d^\mathcal{T}(v_2)=d^\mathcal{T}(v_3)$ as $d^\mathcal{T}(\cdot)$ is an integer. Hence, there is no integer $d^\mathcal{T}(w)$ with $d^\mathcal{T}(v_3)< d^\mathcal{T}(w)< d^\mathcal{T}(v_2)$. Moreover, based on the rule to select $v_3$ by line~9, there is no node $w$ with $d^\mathcal{T}(v_1)< d^\mathcal{T}(w)< d^\mathcal{T}(v_3)$. (ii) $d^\mathcal{T}(v_2)\leq d^\mathcal{T}(v_3)$: Then $d^\mathcal{T}(v_1)\leq d^\mathcal{T}(v_2)\leq d^\mathcal{T}(v_3)$. If there exists $d^\mathcal{T}(w)$ with $d^\mathcal{T}(v_2)< d^\mathcal{T}(w)< d^\mathcal{T}(v_3)$, then $v_3$ is not chosen according to line~9. If there exists $d^\mathcal{T}(w)$ with $d^\mathcal{T}(v_1)< d^\mathcal{T}(w)< d^\mathcal{T}(v_2)$, then we should select this $w$ as $v_3$, thus contradicting the assumption that $d^\mathcal{T}(v_2)\leq d^\mathcal{T}(v_3)$. Therefore, $d^\mathcal{T}(w)\geq d^\mathcal{T}_{m}$ for $w\in V(\mathcal{T})\setminus \{v_1,v_2,v_3\}$.

\emph{2)} Now we prove the completeness of $\mathcal{S}$. It suffices to show that there exists optimal monitor placement for each triconnected component by using only the nodes in $\mathcal{S}$. There are two possible scenarios: (i) $|\mathcal{T}|=2\ or\ 3$: All nodes in $\mathcal{T}$ are candidates. (ii) $|\mathcal{T}|\geq 4$: (a) If $\mathcal{T}$ is completely identifiable, then $\{v_1,v_2,v_3\}$ can all be selected as monitors and no additional nodes in $\mathcal{T}$ are required to be monitors for completely identifying $\mathcal{T}$; (b) If $\mathcal{T}$ is not completely identifiable, then according to the three categories of triconnected components, $\mathcal{T}$ must be of Category 2.2 or 3 (as $\mathcal{T}$ with $|\mathcal{T}|\geq 4$ is completely identifiable if it is of the other categories). When $\mathcal{T}$ is of Category 2.2, then only links incident to one vantage, e.g., $\mu_3$ in Fig.~\ref{Fig:AlgSketch}(b), in $\mathcal{T}$ are unidentifiable. For such case, as illustrated in Fig.~\ref{Fig:AlgSketch}(b), we characterize the relationship between candidates and vantages $\{\mu_1,\mu_2\}$ by set $\Delta:=\{v_1,v_2,v_3\}\setminus \{\mu_1,\mu_2\}$. Then we select a node in $\Delta$ with the minimum number of adjacent links within $\mathcal{T}$ as a monitor, e.g., $\mu_3$ in Fig.~\ref{Fig:AlgSketch}(b). According to the argument in 1), there is no node $w$ with $d^\mathcal{T}(w)< d^\mathcal{T}_{m}$ for $w\in V(\mathcal{T})\setminus \{v_1,v_2,v_3\}$. Therefore, for all nodes (in $\mathcal{T}\setminus \{\mu_1,\mu_2\}$) capable of ensuring $\mathcal{T}$ to be of Category 2.2, the one with the minimum number of adjacent links (actually can be chosen within $\Delta$) is selected as a monitor, i.e., the optimal case for $\mathcal{T}$ of Category 2.2. Now consider the case that $\mathcal{T}$ with $|\mathcal{T}|\geq 4$ is of Category 3. In this case, if two monitors can be selected from $\mathcal{T}$, then $v_1$ and $v_2$ are chosen to be the two vantages in $\mathcal{T}$; therefore, the number of unidentifiable links in $\mathcal{T}$ is minimized (by lines~3 and 5), i.e., $|\mathcal{U}_{v_1 v_2}|$ (or $|\mathcal{U}_{v_1 v_2}|-1$ when there exists a direct link $v_1v_2$ and $v_1v_2$ is identifiable by Algorithm~B). If there exists a vantage $\mu_0$ and only one monitor can be chosen from $\mathcal{T}$, then $v=\operatorname*{arg\min}_{v\in \{v_1,v_2,v_3\}\setminus \mu_0}d(v)$ is selected as a monitor. Therefore, the generated candidate set $\mathcal{S}$ is complete.
\hfill$\blacksquare$

\subsection{Proof of Theorem~\ref{theorem:optimal2VC}}
We prove it by contradiction.

\emph{1)} We prove $\mathcal{O}^*_{\kappa}\subseteq \mathcal{M}$.

(i) When $|\mathcal{M}|=3$, then it is trivial that $\mathcal{O}^*_{3}= \mathcal{M}$.

\begin{figure}[tb]
\centering
\includegraphics[width=2.5in]{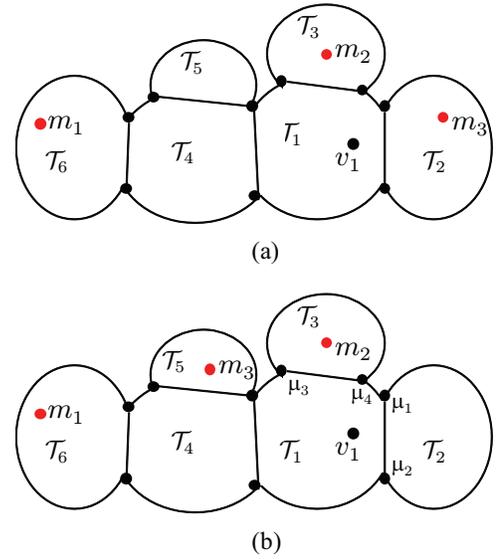}
\vspace{-.5em}
\caption{Optimal monitor placement in 2-vertex-connected networks.}\label{fig:Optimal2VC}
\vspace{-.5em}
\end{figure}

(ii) When $|\mathcal{M}|>3$, as $\kappa\geq 3$ and $\mathcal{G}$ is 2-vertex-connected, each triconnected component involves at least 2 independent vantages. Therefore, all links in triconnected components (except for some links in triangles of Category 2) with 3 or more vantages are identifiable, and only links incident to the two vantages in the triconnected components of Category 3 are unidentifiable. Suppose $\mathcal{O}^*_{\kappa}$ with $\mathcal{O}^*_{\kappa}\subseteq \mathcal{M}$ does not exist, then it implies that some nodes in $\mathcal{O}^*_{\kappa}$ cannot be selected from $\mathcal{M}$. In this case, at least one node in $\mathcal{O}^*_{\kappa}$ is within a triconnected component with 3 or more separation nodes (see \cite{Ma13IMC}), say $v_1$ in Fig.~\ref{fig:Optimal2VC}, since this location cannot be selected by MMP when $|\mathcal{M}|>3$. If $v_1$ is in the location illustrated as Fig.~\ref{fig:Optimal2VC}(a), i.e., all neighboring triconnected components of $\mathcal{T}_1$ within $\mathcal{G}$ contain monitors, then placing a monitor at $v_1$ does not contribute link identification in $\mathcal{G}$; therefore, placing a monitor at $v_1$ is not the optimal solution. Now suppose $v_1$ is in the location illustrated as Fig.~\ref{fig:Optimal2VC}(b), i.e., at least one neighboring triconnected component ($\mathcal{T}_2$) of $\mathcal{T}_1$ contains no monitors. In this case, placing a monitor at $v_1$ does not contribute link identification in $\mathcal{G}$ either, except for identifying the links in triangles (if $\mathcal{T}_1$ is a triangle). Nevertheless, moving the monitor from $v_1$ to a node\footnote{Theorem~\ref{theorem:optimal2VC} only applies to the case that $\kappa\geq 3$, since the following properties cannot be guaranteed if $\kappa=2$ in $\mathcal{G}$.} (a node other than $\mu_1$ and $\mu_2$) in $\mathcal{T}_2$ can maintain the links which are identifiable when $v_1$ is a monitor. Moreover, links incident to $\mu_1$ and $\mu_2$ in $\mathcal{T}_2$ become identifiable under this new monitor location, which can be selected by MMP (see the algorithm details in \cite{Ma13IMC}), i.e., the new location forms a better placement (which can be selected within $\mathcal{M}$) than $\mathcal{O}^*_\kappa$, contradicting the assumption that $\mathcal{O}^*_{\kappa}$ is the optimal $\kappa$-monitor placement.

Hence, based on (i) and (ii), there exists optimal solution $\mathcal{O}^*_{\kappa}$ with $\mathcal{O}^*_{\kappa}\subseteq \mathcal{M}$.

\emph{2)} Now we prove $\mathcal{O}^*_{\kappa+1}$ can be constructed by $\mathcal{O}^*_{\kappa+1}=\mathcal{O}^*_{\kappa}\cup \{v_{m}\}$, where $v_{m}\in \mathcal{M}$, $\mathcal{O}^*_{\kappa}\subseteq \mathcal{M}$.

\begin{figure}[tb]
\centering
\includegraphics[width=3.2in,height=3.5in]{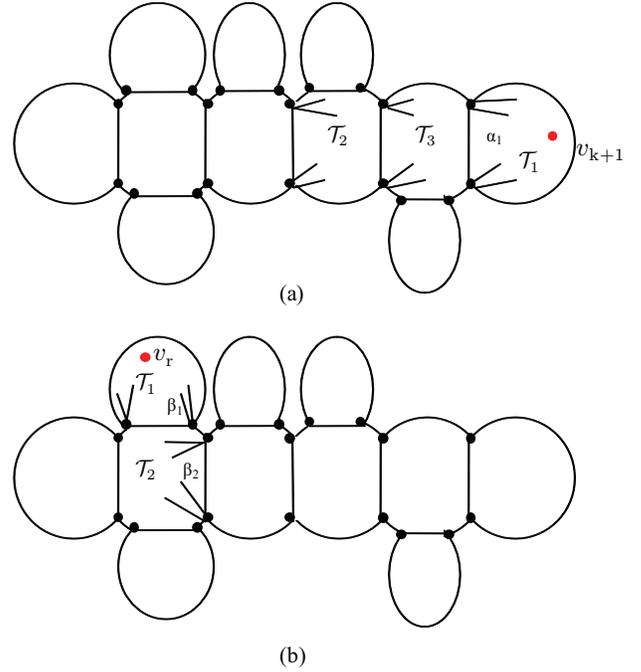}
\vspace{-.5em}
\caption{Nested structure of optimal monitor placement in 2-vertex-connected networks.}\label{fig:optimalProof}
\vspace{-.5em}
\end{figure}

We obtain node set $\mathcal{V}_{\kappa+1}$ by adding node $v_{\kappa+1}$ to set $\mathcal{O}^*_\kappa$ such that $N(\mathcal{V}_{\kappa+1})$ is maximized. Then on top of $\mathcal{O}^*_\kappa$, suppose $|\alpha_1|+|\alpha_2|$ extra identifiable links are achieved by adding node $v_{\kappa+1}$, i.e., $|\alpha_1|+|\alpha_2| = N(\mathcal{V}_{\kappa+1})-N(\mathcal{O}^*_\kappa)$, where $\alpha_1$ is the set of effective exterior links in the triconnected component (e.g., $\mathcal{T}_1$ in Fig.~\ref{fig:optimalProof}(a)) involving $v_{\kappa+1}$ when nodes in $\mathcal{O}^*_\kappa$ are employed as monitors, and $\alpha_2$ is the set of all other identifiable links (e.g., effective exterior links in $\mathcal{T}_2$ and $\mathcal{T}_3$ in Fig.~\ref{fig:optimalProof}(a)) determined by adding $v_{\kappa+1}$. Now suppose $N(\mathcal{V}_{\kappa+1})<N(\mathcal{O}^*_{\kappa+1})$, then moving nodes in $\mathcal{V}_{\kappa+1}$ to specific locations, i.e., $\mathcal{V}'_{\kappa+1}$, can get $N(\mathcal{V}'_{\kappa+1})>N(\mathcal{V}_{\kappa+1})$. For $\mathcal{V}'_{\kappa+1}$, there are two possible cases.

(i) $v_{\kappa+1}\in\mathcal{V}'_{\kappa+1}$: In $\mathcal{V}'_{\kappa+1}$, let $\mathcal{V}_{\kappa}:=\mathcal{V}'_{\kappa+1}\setminus v_{\kappa+1}$. Since $\kappa\geq 3$, we have 3 cases for $\mathcal{V}_{\kappa}$, denoted by $\mathcal{V}^{(1)}_{\kappa}$, $\mathcal{V}^{(2)}_{\kappa}$, and $\mathcal{V}^{(3)}_{\kappa}$. If $\mathcal{V}^{(1)}_{\kappa}$ can determine the identification of all links in $\alpha_2$, then
\begin{equation}\label{eq:withVcase1}
    N(\mathcal{V}'_{\kappa+1})=N(\mathcal{V}^{(1)}_{\kappa})+|\alpha_1|;
\end{equation}
if $\mathcal{V}^{(2)}_{\kappa}$ can only determine the identification of some links in $\alpha_2$, then
\begin{equation}\label{eq:withVcase2}
    N(\mathcal{V}'_{\kappa+1})=N(\mathcal{V}^{(2)}_{\kappa})+|\alpha_i|,
\end{equation}
where $|\alpha_i|$ is an integer with $|\alpha_1|\leq |\alpha_i| \leq |\alpha_1|+|\alpha_2|$; finally, if $\mathcal{V}^{(3)}_{\kappa}$ cannot determine the identification of any links in $\alpha_1$ and $\alpha_2$, then we have
\begin{equation}\label{eq:withVcase3}
    N(\mathcal{V}'_{\kappa+1})=N(\mathcal{V}^{(3)}_{\kappa})+|\alpha_1|+|\alpha_2|.
\end{equation}
We know that $N(\mathcal{V}^{(1)}_{\kappa})\leq N(\mathcal{O}^*_\kappa)$, $N(\mathcal{V}^{(2)}_{\kappa})\leq N(\mathcal{O}^*_\kappa)$, and $N(\mathcal{V}^{(3)}_{\kappa})\leq N(\mathcal{O}^*_\kappa)$. Moreover, according to the way of getting $\mathcal{V}_{\kappa+1}$, it is achievable for $N(\mathcal{V}^{(3)}_{\kappa})$ to be $N(\mathcal{O}^*_\kappa)$. Therefore, the best case for $\mathcal{V}'_{\kappa+1}$ is that $N(\mathcal{V}^{(3)}_{\kappa})=N(\mathcal{O}^*_\kappa)$ in (\ref{eq:withVcase3}), which implies $N(\mathcal{V}'_{\kappa+1})=N(\mathcal{V}_{\kappa+1})$, contradicting the assumption that $N(\mathcal{V}'_{\kappa+1})>N(\mathcal{V}_{\kappa+1})$.

(ii) $v_{\kappa+1}\notin\mathcal{V}'_{\kappa+1}$: In this case, there exists node $v_r$ with $v_r\notin\mathcal{O}^*_{\kappa}$ in $\mathcal{V}'_{\kappa+1}$. Accordingly, $\mathcal{V}'_{\kappa+1}$ can be written in the form of $\mathcal{V}'_{\kappa+1}=\mathcal{V}_{\kappa}\cup v_r$. Then on top of $\mathcal{V}_\kappa$, suppose $|\beta_1|+|\beta_2|$ extra identifiable links are achieved by adding node $v_{r}$, where $\beta_1$ is the set of effective exterior links in the triconnected component (e.g., $\mathcal{T}_1$ in Fig.~\ref{fig:optimalProof}(b)) involving $v_{r}$ when nodes in $\mathcal{V}_\kappa$ are employed as monitors, and $\beta_2$ is the set of all other identifiable links (e.g., effective exterior links in $\mathcal{T}_2$ in Fig.~\ref{fig:optimalProof}(b)) determined by adding $v_{r}$. Then moving nodes in $\mathcal{V}_{\kappa}$ to other locations, i.e., $\mathcal{V}'_{\kappa}$, there are three possible cases, denoted by $\mathcal{V}^{(1)}_{\kappa}$, $\mathcal{V}^{(2)}_{\kappa}$, and $\mathcal{V}^{(3)}_{\kappa}$. If $\mathcal{V}^{(1)}_{\kappa}$ can determine the identification of all links in $\beta_2$, then
\begin{equation}\label{eq:withNoVcase1}
    N(\mathcal{V}'_{\kappa+1})=N(\mathcal{V}^{(1)}_{\kappa})+|\beta_1|;
\end{equation}
if $\mathcal{V}^{(2)}_{\kappa}$ can only determine the identification of some links in $\beta_2$, then
\begin{equation}\label{eq:withNoVcase2}
    N(\mathcal{V}'_{\kappa+1})=N(\mathcal{V}^{(2)}_{\kappa})+|\beta_i|,
\end{equation}
where $|\beta_i|$ is an integer with $|\beta_1|\leq |\beta_i| \leq |\beta_1|+|\beta_2|$; finally, if $\mathcal{V}^{(3)}_{\kappa}$ cannot determine the identification of any links in $\beta_1$ and $\beta_2$, then we have
\begin{equation}\label{eq:withNoVcase3}
    N(\mathcal{V}'_{\kappa+1})=N(\mathcal{V}^{(3)}_{\kappa})+|\beta_1|+|\beta_2|.
\end{equation}
Note that the correctness of (\ref{eq:withNoVcase1})--(\ref{eq:withNoVcase3}) are ensured by the fact that $\kappa\geq 3$. Following the similar argument in (i), since $v_r\notin\mathcal{O}^*_{\kappa}$, it is achievable for $N(\mathcal{V}^{(3)}_{\kappa})$ to be $N(\mathcal{O}^*_\kappa)$ in (\ref{eq:withNoVcase3}), which implies $\mathcal{V}'_{\kappa+1}=\mathcal{O}^*_{\kappa}\cup v_r$. We know $v_r\neq v_{\kappa+1}$, $v_{\kappa+1}\notin \mathcal{O}^*_{\kappa}$, and  $N(\mathcal{O}^*_{\kappa}\cup v_r)\leq N(\mathcal{O}^*_{\kappa}\cup v_{\kappa+1})$; therefore, even for the best case of $\mathcal{V}'_{\kappa+1}$ with $v_{\kappa+1}\notin\mathcal{V}'_{\kappa+1}$, it cannot achieve a larger number of identifiable links than that determined by $\mathcal{O}^*_\kappa\cup v_{\kappa+1}$. Thus, no matter how we move the nodes in $\mathcal{V}_{\kappa+1}$ to other locations, it is impossible to find a monitor placement achieving a larger number of identifiable links. Therefore, $\mathcal{V}_{\kappa+1}=\mathcal{O}^*_{\kappa+1}$.

Consequently, the optimal $(\kappa+1)$-monitor ($\kappa+1\leq |\mathcal{M}|$) placement $\mathcal{O}^*_{\kappa+1}$ can be constructed by $\mathcal{O}^*_{\kappa+1}=\mathcal{O}^*_{\kappa}\cup \{v_{m}\}$, where $v_{m}={\operatorname*{arg\max}}_{v}N(\mathcal{O}^*_{\kappa}\cup \{v\})$ over $v \in \mathcal{M}\setminus \mathcal{O}^*_{\kappa}$.
\hfill$\blacksquare$

\section{Performance Evaluation of GMMP}
Besides the main simulation results in \cite{MaInfocom14sub}, we also evaluate algorithm performance in terms of algorithm running time. Using the same set of configurations at that in the main simulation results of \cite{MaInfocom14sub}, the running times in Fig.~\ref{fig:simRunningTime} shows that GMMP is also faster than RMP-$V$ and RMP-$\mathcal{S}$.

\addtocounter{footnote}{0}
{
\begin{figure}[tb]
\centering
\vspace{-.8em}
\includegraphics[width=3.8in]{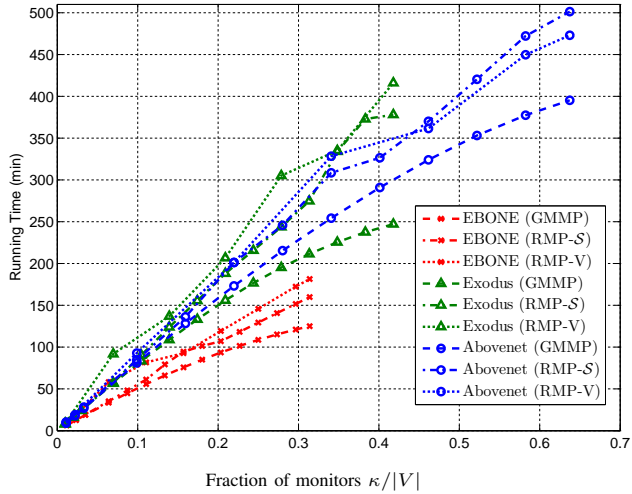}
      \vspace{-2em}
      \scriptsize \\ \ \ \ \ \ Fraction of monitors $\kappa/|V|$\normalsize
      \vspace{-.4em}
\caption[blah]%
{Algorithm running time in ISP networks\footnotemark.} \label{fig:simRunningTime}
\vspace{-.1em}
\end{figure}
\footnotetext{The running time for an exhaustive search is significantly larger (over $48$ hours for each $\kappa$), and is thus omitted in Fig.~\ref{fig:simRunningTime}.}
}

\bibliographystyle{IEEEtran}
\bibliography{mybibSimplified}
\end{document}